\documentclass[aps,pre,onecolumn,superscriptaddress,groupedaddress]{revtex4}  %
\usepackage{amssymb}   
\usepackage{amsmath}
\usepackage{mathtools}
\usepackage{multirow}
\usepackage{makecell}
\usepackage{graphicx}  
\usepackage{xcolor}
\usepackage{soul}
\newcommand{\cmt}[1]{\textcolor{black}{#1}}
\newcommand{\cmtt}[1]{\textcolor{black}{#1}}
\newcommand{\cmttt}[1]{\textcolor{black}{#1}}

\begin{document}


\title{Induced Percolation on Networked Systems}
\author{Jiarong Xie}\thanks{These two authors contributed equally}
\affiliation{School of Information Science and Technology, Sun Yat-sen University, Guangzhou 510006, China}
\author{Xiangrong Wang}\thanks{These two authors contributed equally}
\affiliation{Institute of Future Networks, Southern University of Science and Technology, Shenzhen, China}
\affiliation{Research Center of Networks and Communications, Peng Cheng Laboratory, Shenzhen, China}
\author{Ling Feng}
\affiliation{Institute of High Performance Computing, A*STAR, 138632 Singapore}
\affiliation{Department of Physics, National University of Singapore, Singapore 117551}
\author{Jin-Hua Zhao}
\affiliation{Guangdong Provincial Key Laboratory of Nuclear Science, Institute of Quantum Matter, South China Normal University, Guangzhou 510006, China}
\author{Yamir Moreno}
\affiliation{Institute for Biocomputation and Physics of Complex Systems,University of Zaragoza, 50018, Zaragoza, Spain}
\affiliation{Department of Theoretical Physics, University of Zaragoza, 50018, Zaragoza, Spain}
\affiliation{ISI Foundation, 10126, Torino, Italy}
\author{Yanqing Hu}\email{yanqing.hu.sc@gmail.com}
\affiliation{School of Information Science and Technology, Sun Yat-sen University, Guangzhou 510006, China}
\affiliation{Southern Marine Science and Engineering Guangdong Laboratory, Zhuhai 519082, China}

\begin{abstract}
Percolation theory has been widely used to study phase transitions in complex networked systems. It has also successfully explained several macroscopic phenomena across different fields. Yet, the existent theoretical framework for percolation places the focus on the direct interactions among the system's components, while recent empirical observations have shown that indirect interactions are common in many systems like ecological and social networks, among others. Here, we propose a new percolation framework that accounts for indirect interactions, which allows to generalize the current theoretical body and understand the role of the underlying indirect influence of the components of a networked system on its macroscopic behavior. We report a rich phenomenology in which first-order, second-order or hybrid phase transitions are possible depending on whether the links of the substrate network are directed, undirected or a mix, respectively. We also present an analytical framework to characterize the proposed induced percolation, paving the way to further understand network dynamics with indirect interactions.
\end{abstract}
\maketitle


Percolation theory  \cite{stauffer2018introduction} is no doubt one of the most prominent frameworks within statistical physics. Initially developed  \cite{flory1941molecular,stockmayer1943theory} to explain the chemical formation of large macromolecules, it has been recently used to study various dynamical process in complex networks \cite{Newman.Strogatz.Watts-PRE-2001,Callaway.etal-PRL-2000,Cohen.etal-PRL-2000,Watts-PNAS-2002,Dorogovtsev2008critical,castellano2009statistical}. Examples include the use of bond percolation \cite{castellano2009statistical,hu2018local}  to study the wide spread of rumors over online social media and outbreaks of infectious diseases on structured populations. Site percolation \cite{Callaway.etal-PRL-2000,Newman.Strogatz.Watts-PRE-2001,bashan2013extreme} has been employed to study the cascading failures of infrastructure networks \cite{Parshani.Buldyrev.Havlin-PRL-2010,Cohen.etal-PRL-2000,buldyrev2010catastrophic,gao2012networks,brummitt2012suppressing}, and the resilience of protein-protein interaction networks \cite{albert2000error}. Likewise, bootstrap percolation \cite{baxter2010bootstrap}, $k$-core \cite{Dorogovtsev.etal-PRL-2006,Baxter.etal-PRX-2015,Zhao.Zhou.-NatCommun-2013} and linear threshold percolation \cite{Granovetter-AmJSocial-1978,Watts-PNAS-2002,kempe2003maximizing,Morone.Makse-Nature-2015} have enabled the study of the spreading of behaviors over social networks. Finally, so-called explosive percolation \cite{achlioptas2009explosive} has allowed a better characterization of systems' structural transitions when they are growing or can adapt, whereas core percolation \cite{Bauer.Golinelli-EPJB-2001,Liu.Csoka.Zhou.Posfai-PRL-2012} has contributed significantly to get insights into NP problems. Common to all these percolation models is that they have successfully described various important dynamical phenomena by considering different {\it direct} interactions \cite{Dorogovtsev2008critical,castellano2009statistical,Cohen.Havlin-2010} over network nodes, in particular, they have captured the behavior of networked systems as given by phase transitions \cite{newman2018networks,Newman.Strogatz.Watts-PRE-2001,Dorogovtsev2008critical,castellano2009statistical,Cohen.Havlin-2010}. 

Our study is motivated by recent evidences that have shown that there are many systems in which {\it indirect} interactions play a major role on their dynamics \cite{Christakis.Fowler-NEJM-2007,Fowler.Christakis-PNAS-2010,guimaraes2017indirect,ohgushi2012trait,lehn1993supramolecular,gierschner2009excitonic}. Such underlying indirect interactions might have important implications not only on the dynamics of the system, but also on the evolution and the emergence of network structures. For example, Christakis and Fowler \cite{Christakis.Fowler-NEJM-2007,Fowler.Christakis-PNAS-2010} found that for many social behaviors that can spread, such as drug \cite{rudolph2013individual} and alcohol addictions \cite{rosenquist2010spread} and obesity \cite{Christakis.Fowler-NEJM-2007}, an individual can span its influence to their friends around three degrees of separation (friend of a friend's friend). This phenomenon is also widely known as ``three degrees of influence'' in social science, which is an important behavioral influence mechanism that has a large impact on public health. In ecological networks, Guimar{\~a}es Jr et al. \cite{guimaraes2017indirect,ohgushi2012trait} discovered in 2017 that indirect effects contributed strongly to the trait coevolution among reciprocal species, which can alter environmental selection and promote the evolution of species. A final example is given by chemical molecular networks, where neglecting indirect molecular interactions significantly underestimates some kind of coupling in molecular assemblies \cite{lehn1993supramolecular,gierschner2009excitonic}.

Notwithstanding the previous evidences, up to date there has been no percolation-based theoretical model that is suited to describe indirect influence, nor the consequences of having such interactions for the macroscopic behavior of the system. This is because, as mentioned before, no matter whether the interactions are encoded in regular or complex networks, existing models of percolation are always based on direct relations \cite{Dorogovtsev2008critical,castellano2009statistical,Cohen.Havlin-2010} among nodes. In other words, all of the current models only take into account the existence and the strength of links for directly connected nodes, regardless of any higher-order correlations given by indirect connections with other nodes. Here we fill this gap and propose a new percolation framework to study the impact of indirect interactions on the behavior of the whole system. This mechanism is called induced percolation. 

Our results show that indirect interactions lead to a unique macroscopic behavior characterized by anisotropy and phase transitions. Specifically, we study the most general scenario in which links can be directed and report that varying the links' directness could change the order of the phase transition. This is in direct contrast to previous percolation models, for which the nature of the phase transitions is not affected by the directionality of links. To the best of our knowledge, the phenomenon of directness-related order of the phase transitions only exists in some special cases of core percolation \cite{Liu.Csoka.Zhou.Posfai-PRL-2012}, whereas it is shown to be a generic feature in our indirect interaction model.

\section{Results}\label{section_result}

Induced percolation can be formally defined on directed networks as follows. Let us assume that the state of the nodes is characterized by an integer value, 0 or 1. Initially, we set the state of all nodes in the network to $1$. A node $i$ remains in state $1$ if at least one of its incoming links comes from a node, say $j$, with state $1$, and in turn the node $j$ has at least $m$ other incoming links from nodes that are in state $1$, see Fig.1a for an illustration of the case $m=2$. Otherwise, node $ i $ changes to state $ 0 $ at the next time step. The influence of the $m$ nodes on the node $i$ defines the indirect interactions among them. Under this mechanism, certain nodes will change their states from 1 to 0 at each time step until no more changes are possible. Compared with bond, bootstrap or $k$-core percolation, the fundamental difference of induced percolation is that the current state of a node is affected not only by its nearest neighbors, but also by a number of its next-nearest neighbors.

The mechanism for induced percolation through a network captures the observation that there are behaviors whose influence reaches nodes beyond the first shell. As an illustration of our proposed induced percolation model, consider a social network in which everyone can play Mahjong, which is a game played by 4 people. If the friend of Jack maintains the habit of playing Mahjong, and at the same time, the friend of Jack has two more friends who play Mahjong (corresponding to $m = 2$), then 3 of them (excluding Jack) need a fourth person to make it work. Hence, the friend of Jack will induce/persuade Jack to join the 3, and in turn will make Jack in a sustained state of playing Mahjong. This mechanism corresponds exactly to the $m = 2$  induced percolation on social networks. This sort of process is also very relevant in disease dynamics, when the interest is in tracing back the origin of infected nodes by going backwards in the infection tree: one would like to know what is the likelihood that node $i$ could be infected (state 1) given that its neighbor node $j$ has at least $m$ infected nodes. In other words, what is the probability of observing an infection tree made up by node $i$, node $j$ and $m$ other neighbors of $j$? Though the above two cases exemplify the induced percolation, it is conceivable that the mechanism of direct interactions mediated by indirect neighbors is representative for a large class of propagation or dynamical processes on complex systems.

The main quantity of interest is the outgoing giant component (GOUT) \cmtt{\cite{Dorogovtsev2008critical,Cohen.Havlin-2010,newman2018networks}}. For various types of propagation dynamics on networks, the outgoing giant component corresponds to the largest spreading coverage, and it serves as an indicator for network connectivity under the specified propagation mechanism. The size of the out-going giant component in the above example corresponds to the number of individuals who were infected. Therefore, in induced percolation, the size of the out-going giant component is the order parameter, i.e., the macroscopic quantity that characterizes phase transitions. In addition, we also examine the size distribution of small out-going components. 

\begin{figure*}[h]%
	\centering
	\includegraphics[width=0.9\textwidth]{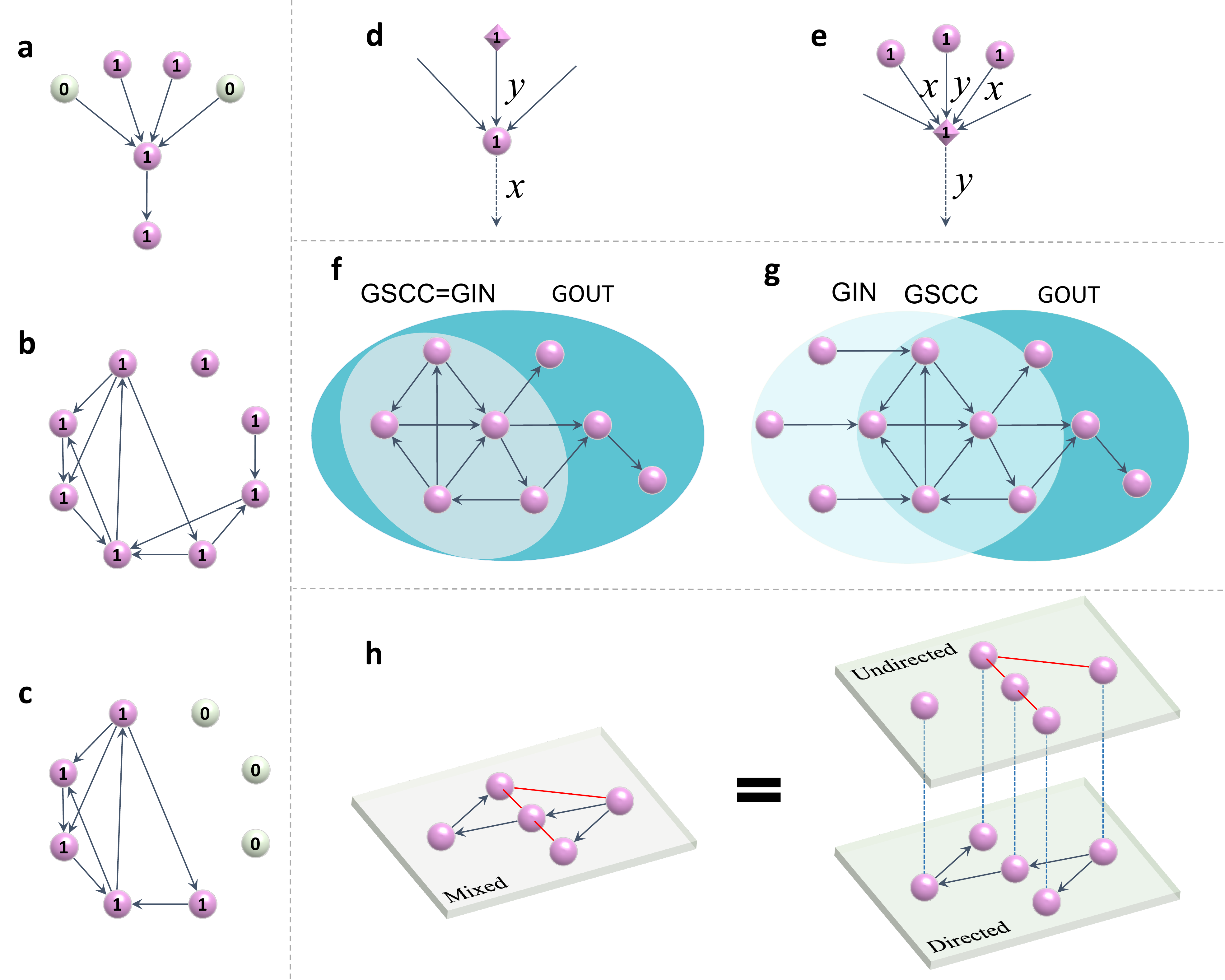}%
	\caption{{\bf Induced percolation on directed networks}. Panel (a) illustrates the proposed mechanism of induced percolation for the case $ m=2 $. In order for a node $ i $ to remain in state $ 1 $, at least one node ($j$) at the other end of an incoming link should be in state $ 1 $. On its turn, $j$ should also have at least $m$ ($=2$ in the example) incoming links from neighbors that are in state $ 1 $. Panel (b) shows a directed graph of $ 8 $ nodes all in state $ 1 $. Panel (c) shows the giant out-going component (GOUT) when the graph on panel (b) is pruned according to the induced percolation rules. Panels (d) and (e) illustrate the variables $x$ and $y$ defined in the main text by equations (\ref{eq:directed_x}-\ref{eq:directed_y}). Panels (f) and (g) show the relationship between the order parameters GSCC, GIN and GOUT, for induced percolation and typical bond percolation processes, respectively. Panels (h) and (i) schematically represent the multilayer representation employed to derive the order parameter $ P_\infty $ when there are directed and undirected links in the substrate network.}
	\label{fig_m_induced_percolation}
\end{figure*}

In undirected networks, each link can be viewed as two directed links with opposite directions. Therefore, induced percolation can be studied on fully directed networks, and then extent the methodology to either undirected (i.e., fully bidirectional) networks or to networks in which there are both bidirectional and unidirectional links. Note that the out-going giant component of undirected networks is the same as the incoming giant component and the strongly-connected giant component. We schematically illustrate the proposed induced percolation mechanism on directed networks in Fig. \ref{fig_m_induced_percolation}, where we also show the order parameter as compared with the one typically used in bond percolation. Similar diagrams for undirected and mixed networks can be found in the Methods and Supplementary Information.

The phase transition that characterizes the induced percolation process can be analytically studied on random networks. The class of random directed networks is constructed by independently connecting two arbitrary nodes with a directed link with a fixed probability. The network can be described by the joint degree distribution $ P\left(k_{\text{in}},k_{\text{out}}\right) $, which is the probability that a randomly selected node has out-degree $k_{\text{out}}$ and in-degree $k_{\text{in}}$. For random directed networks, the size of GOUT is derived through the following recursive equations. We first define two recursive variables $ x $ and $ y $ (see \cmttt{Fig.} \ref{fig_m_induced_percolation}d,e): $ x $ represents the probability that when selecting at random a directed link, the node at the origin of the link is active (in state $ 1 $), whereas $y$ represents the probability that a link enables its end node to be in an active state. According to the definitions of $ x $ and $ y $, we have

\begin{equation}\label{eq:directed_x}
x = \sum _{k_{\text{in}},k_{\text{out}}}^{+\infty}\frac{k_{\text{out}}P\left(k_{\text{in}},k_{\text{out}}\right)}{\left<k\right>} \left[ 1 - (1 - y)^{k_{\text{in}}} \right].
\end{equation}
where $ 1- (1-y)^{k_{\text{in}}} $ is the probability that, for a node with an incoming degree $ k_{\text{in}} $, at least one of the $k_{\text{in}}  $ incoming neighbors is active (i.e., in state 1). $ \frac{k_{\text{out}}P\left(k_{\text{in}},k_{\text{out}}\right)}{\left<k\right>} $ is the excess degree distribution \cite{Cohen.Havlin-2010,newman2018networks} for the node at the origin of an arbitrary directed link. This is because the likelihood of a node being the origin of a randomly chosen directed link is proportional to the node's out-degree.

Calculating the probability $ y $ is a little more involved. The definition of the induced percolation process (see Figure \ref{fig_m_induced_percolation}) implies that even if the starting node of a directed link is active (which happens with probability $x$), it is not guaranteed that the end node of this directed link remains active (which happens with probability $y$). However, if the starting node of this directed link is itself active, and at the same time at least $ m $ neighbors pointing to the starting node are active, then this directed link can keep its end node active. Conversely, if a directed link can keep the node it points to active (corresponding to $ y $), then the starting node of this directed link must be active (corresponding to $ x $). Therefore, it must hold $ x> y $ when $ m> 1 $ ($ x = y $ when $ m = 1 $ which corresponds to bond percolation). The above analysis yields the expression of $ y $ as:
\begin{equation}\label{eq:directed_y}
	y = \sum _{k_{\text{in}},k_{\text{out}}}^{+\infty}  \frac{k_{\text{out}}P\left(k_{\text{in}},k_{\text{out}}\right)}{\left<k\right>} \sum _{s = m}^{k_{\text{in}}} {k_{\text{in}} \choose s}x^s(1-x)^{k_{\text{in}} - s}\left[ 1-\left(1-\frac{y}{x}\right)^s \right] 
\end{equation}
where $ {k_{\text{in}} \choose s} x^s(1-x)^{k_{\text{in}} - s} $ gives the probability that for a node of in-coming degree $ k_{\text{in}} $, $ s $ out of $ k_{\text{in}} $ neighbors are active (in state 1), while $1- (1-y / x)^s $ is the probability that at least 1 out of the $s$ active incoming neighbors keeps this node active (in state 1).

Using equations (1) and (2) we can solve for $ x $ and $ y $, which can then be used to calculate the order parameter $ P_\infty $ of the out-going component size from the equation:
\begin{equation}\label{eq:directed_percolation_prob}
P_{\infty} = \sum _{k_{\text{in}},k_{\text{out}}}^{+ \infty}  P\left(k_{\text{in}}, k_{\text{out}}\right) [1-\left(1-y\right)^{k_{\text{in}}}]
\end{equation}
Here $ P_\infty $ is equivalent to the probability that a randomly chosen node has at least one incoming node to keep it active. One interesting finding worth highlighting is that the GSCC coincides with the GIN for the induced percolation process on directed networks, which is not the case for classical percolation models (see Fig. \ref{fig_m_induced_percolation}f,g). The theoretical analysis of the order parameter $ P_{\infty} $ on undirected networks is illustrated in the Methods section. We also note that the analysis of $ P_{\infty} $ on mixed networks can be done by mapping the structure to a multilayer network, see Fig \ref{fig_m_induced_percolation}h,i and more details in the Supplementary Information. 

\begin{figure*}[h]
	\centering
	\includegraphics[width=1\textwidth]{{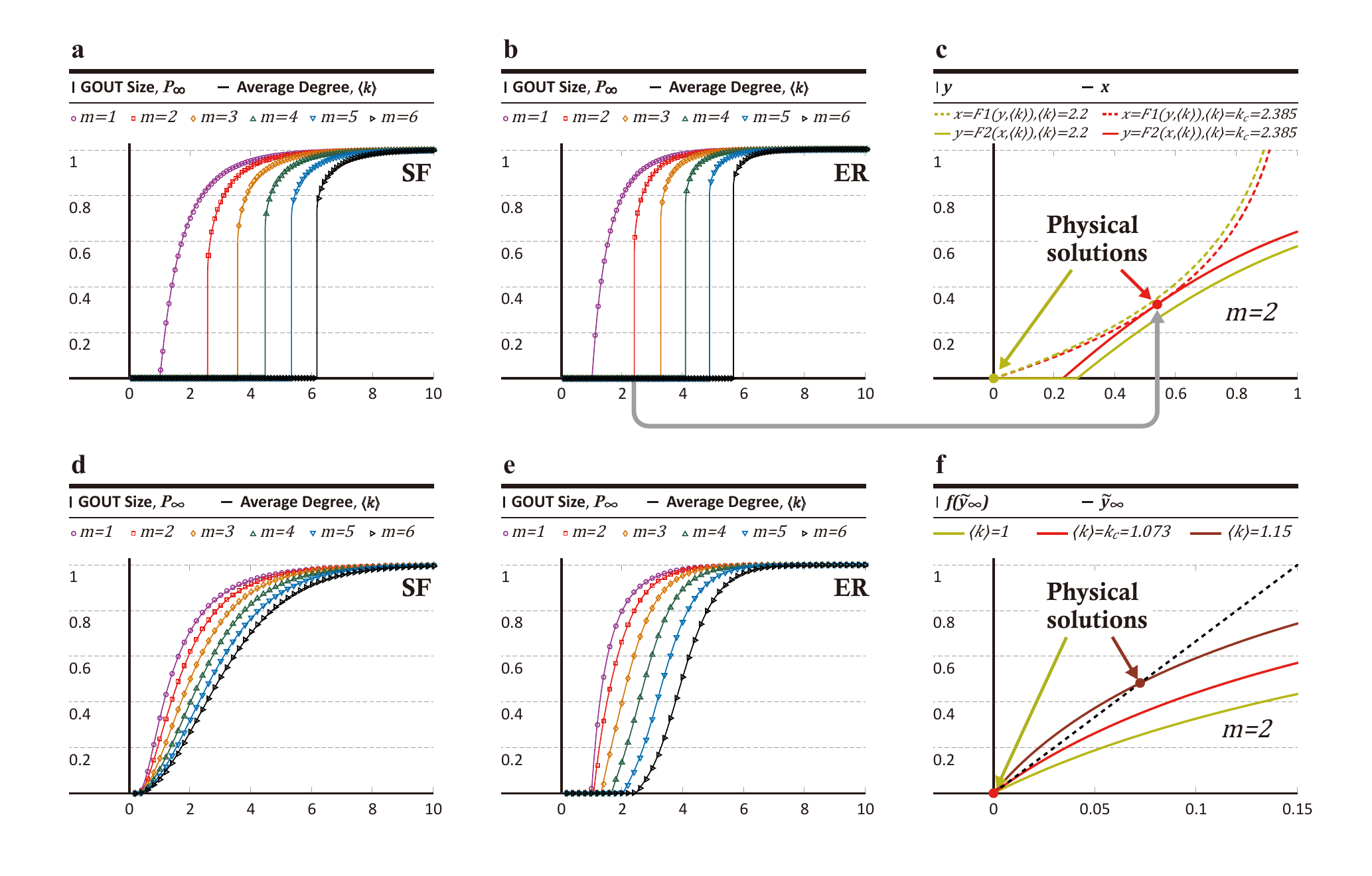}}%
	\caption{{\bf Giant out-going component for induced percolation on directed and undirected random networks}. Panels (a-b) show the order parameter GOUT for induced percolation ($m=2,\ \ldots,\ 6 $) on directed Scale-Free (SF) and Erd\H{o}s-R\'enyi (ER) networks as a function of the average degree $ \left<k\right> $. Results are compared with the behavior of the same order parameter for bond percolation (equivalent to setting $m=1$). Panel (c) shows the graphical solution of Eq. (\ref{eq:directed_y}) for induced percolation ($m=2$) on directed ER graphs, where $ k_c $ is the critical average degree at which a first order phase transition takes place. Panels (d) and (e) show results for undirected networks, where as the graphical solution shown in panel (f) is derived from Eq.(\ref{eq:undi_y_inf}) (see Methods) for induced percolation ($ m=2 $) on undirected ER graphs. Directed SF networks are generated setting the exponents of the incoming and outgoing degree distributions to $ \gamma_{\text{in}}=3.5 $ and $ \gamma_{\text{out}}=3.0 $, respectively.}
	\label{fig_order_parameter_undirected_directed}
\end{figure*}

Theoretical analyses allow to show that the type or order of the phase transition depends on the directionality of the links for the same network connectivity pattern, i.e., the phase transition is anisotropic in nature. On directed networks, when $m>1$ ($m=1$ is the case of typical bond percolation), induced percolation shows discontinuous (first-order) phase transitions (Fig. \ref{fig_order_parameter_undirected_directed}a,b,c and Fig. \ref{fig:realNet}a,b for real world networks). Yet, on undirected networks, the same percolation process always leads to continuous (second-order) phase transitions (see Fig. \ref{fig_order_parameter_undirected_directed}d,e,f and Fig. \ref{fig:realNet}a,b for real world networks). These results are in sharp contrast with previous percolation models on networks, for which it has never been found that the directness of network links fundamentally alters the type of phase transitions. This means that previously studied types of percolation models, might have significantly underestimated the effects of asymmetry in link directions on the system's macroscopic behavior. An important implication of this observation is that abrupt transitions in complex systems like ecological and social networks might be way more likely to occur than previously anticipated by existent percolation models.

\begin{figure*}[h]%
	\centering
	\includegraphics[width=1\textwidth]{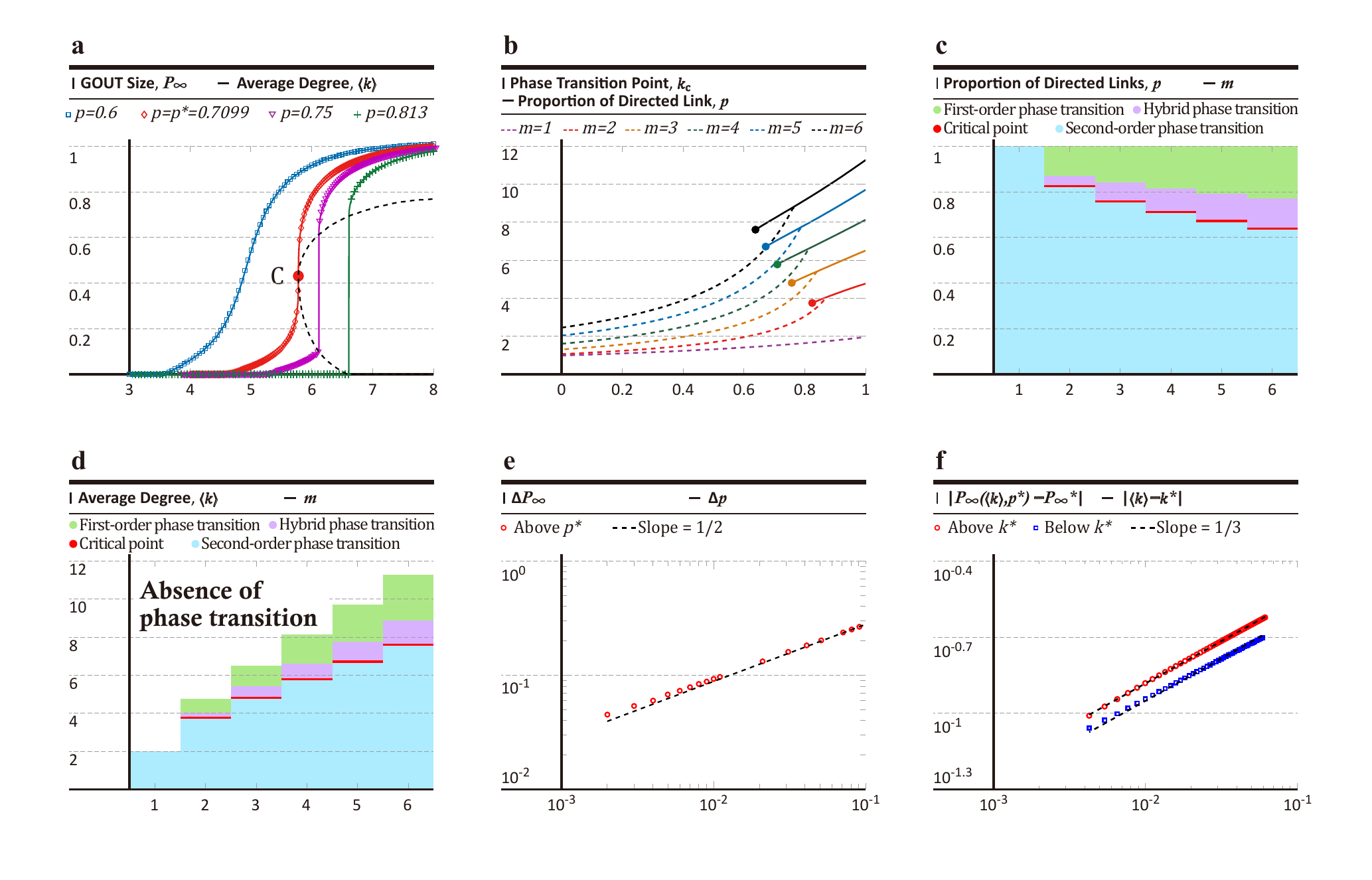}%
	\caption{{\bf Phase transitions and critical behaviors of induced percolation on mixed networks}. In panel (a), we show theoretical and numerical results for GOUT as a function of the average degree $ \left<k\right> $ when the fraction of directed links is varied. The point $C$ denotes the critical point at which coexistence of second- and first-order phase transitions occurs for the first time. The curved dotted line represents the value of GOUT before and after the first-order phase transition. Panel (b) shows the values of the critical points in the parameter space made up by the average degree and the percentage of directed links: the dotted line describes the critical value at which a second-order phase transition occurs, while the solid line corresponds to the first-order phase transition. Dots correspond to critical points, $C$. Panel (c) represents the types of phase transitions that can be observed in the $m-p$ plane. Blue, purple, and green colors bound the area in which second-order, hybrid, and first-order phase transitions exist, respectively. The red boundary lines between the blue and the purple areas correspond to the critical points $C$. When the parameters are such that they lay on the red line, the behavior of GOUT corresponds to the green line marked with point $C$ in panel (a). Panel (d) shows the types of phase transitions shown in (c) but in the $m-\left<k\right> $ plane. Panel (e) presents results of the jump size, $ \Delta P_{\infty} $, as a function of $ \Delta p = p-p^* $ when the critical point $C$ is approached either from below or from above. Panel (f) depicts the change of $P_{\infty}$ near the critical point $ k^* $ as a function of $ \left<k\right>-k^* $, when fixing $ p=p^* $. The mixed network is generated by assigning a percentage $p$ of directed links to an undirected ER network with an average degree $ \left<k\right> $ and consists of $10^6$ nodes.}
	\label{fig_phase_transition_mixed}%
\end{figure*}

The anisotropy induced by the directionality of the links leads to a rich and complex behavior when the network is composed of a mixture of directed and undirected links. Specifically, a hybrid phase transition emerges with the presence of a certain amount of directed links. Figures \ref{fig_phase_transition_mixed}a, b and Fig. \ref{fig:realNet}c show that by increasing the fraction $p$ of directed links in the network, the order parameter GOUT evolves, as the average degree $\left<k\right>$ increases, from a continuous transition to a hybrid phase transition where both continuous and discontinuous transition exist, to a first order transition for larger values of $\left<k\right>$. In addition, in the region where the hybrid phase transition is observed, several quantities follow a set of scaling relations with critical exponents that are in line with Landau's mean-field theory. We label the critical hybrid point where the hybrid transition first appears as point $C\left(k^*,\ p^*\right) $ in Figure \ref{fig_phase_transition_mixed}a. We find a set of scaling relations connecting GOUT to other quantities near $C$ that are predicted by Landau's mean field theory: within the hybrid transition, the jump height of GOUT, $\Delta P_\infty(p^*+\Delta p) \coloneqq \lim_{\left<k\right> \rightarrow k_c^{+}}P_\infty(\left<k\right>,p^*+\Delta p)- \lim_{\left<k\right> \rightarrow k_c^{-}}P_\infty (\left<k\right>,p^*+\Delta p) $ where $ k_c $ is the critical point at which the first order transition occurs, follows a scaling function of $ \Delta p$ with the critical exponent $\eta = 1/2$ (Fig. \ref{fig_phase_transition_mixed}e)
\cmttt{
\begin{equation}
\Delta P_\infty(p^*+\Delta p)  \sim \left(\Delta p\right)^\frac{1}{2}.
\end{equation}}
The same critical exponent holds for the jump height as a scaling function of $ \left<k\right> -k^* $ as shown in the Supplementary Information. When fixing $ p $ at $ p^* $ and \cmttt{varying $ \left<k\right> $} in the vicinity of $ k^* $, the size deviation of GOUT is quantified by the following scaling function of $ \left<k\right>-k ^ * $ with critical exponent $ \theta = 1/3 $ (Fig. \ref{fig_phase_transition_mixed}f), reached from both below and above, 
\begin{equation}
\left| P_\infty(\left<k\right>,p^*)- P_\infty^* (k^*,p^*)\right|  \sim \left|\left<k\right>-k^* \right|^{\frac{1}{3}}.
\end{equation}
We note that Baxter et al. also find these two critical exponents in $k$-core percolation \cite{Baxter.etal-PRX-2015}.

\begin{figure*}[h]
	\centering
	\includegraphics[width=1\textwidth]{{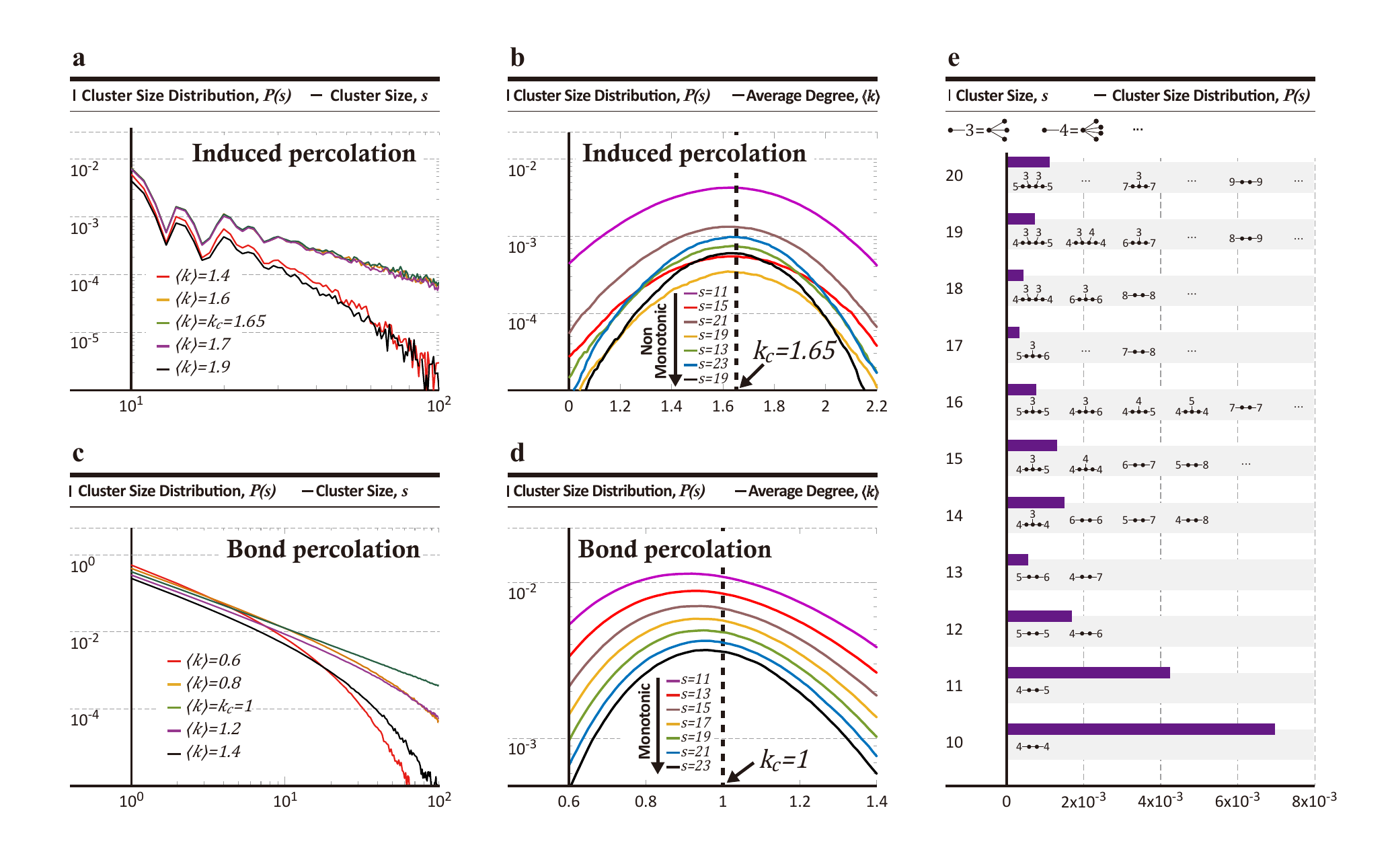}}%
	\caption{{\bf Size distribution,} $P(s)${\bf, of small clusters at the critical point of induced percolation (}$m=4${\bf) on undirected networks}. In (a) we show the size distribution, $P(s)$, which exhibits a fluctuating behavior especially for small sizes. Panel (b) plots the same distribution $P(s)$ but as a function of the average degree $ \left<k\right> $, showing an unambiguous non-monotonic decrease of the size distribution. Panels (c) and (d) depict the monotonous power-law decay of the cluster size distribution in the limit of classical bond percolation. Finally, panel (e) displays the cluster size distribution at the critical point $ k_c=1.65 $, also showing the structure of each cluster. Results are averaged over $10^3$ independent realizations of undirected ER networks (of size $10^6$ nodes). As it can bee clearly seen, the critical behaviors of induced and classical percolation processes differ.}
	\label{fig_small_cluster_distribution}%
\end{figure*}

\begin{figure*}
	\includegraphics[width=1\textwidth]{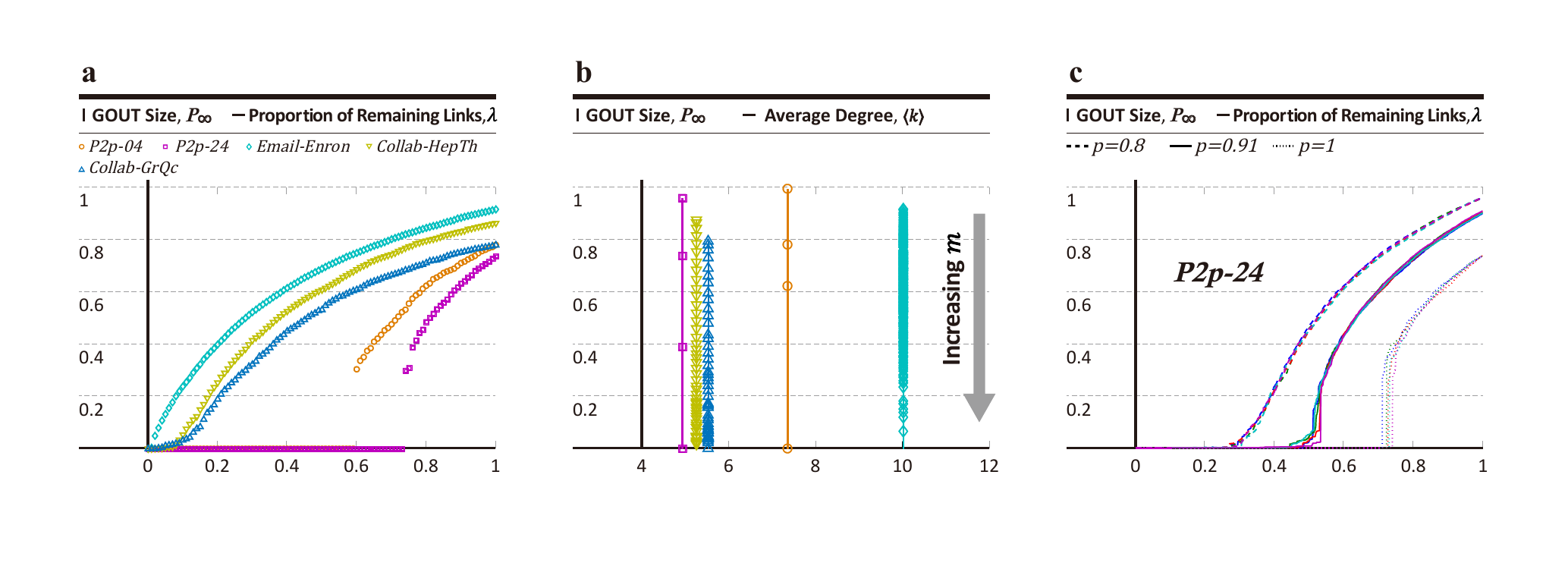}
	\caption{{\bf Order parameter GOUT for induced percolation on top of real-world directed, undirected and mixed networks}. Panel (a) shows GOUT as a function of the proportion, $\lambda$, of remaining links. Each point of GOUT is computed for an induced percolation process with $m=2$ on real-world networks after a fraction of $1-\lambda $ links has been randomly removed. For directed \emph{P2p-04, P2p-24} networks, GOUT shows a discontinuous transition, while for undirected \emph{Email-Enron, Collab-HepTh and Collab-GrQc} networks, GOUT shows a continuous transition. Panel (b) shows results for GOUT for various values of the parameter $m$ on the original real-world networks without link removal ($ \lambda=1 $). When increasing the induced percolation parameter $m$ from $m=1$, GOUT experiences abrupt changes on directed networks and relatively small ones on undirected networks. Finally, panel (c) depicts results of the behavior of GOUT as a function of the remaining fraction of links, $ \lambda $, for real-world mixed networks and $m=2$. Mixed networks are obtained converting, in the \emph{the P2p-24 directed networks}, a proportion ($1-p$) of directed links into undirected links. As observed for synthetic networks, gradually increasing the proportion of directed links $p$ leads first to a continuous, then to a hybrid and finally to a discontinuous transition. More detailed information of the real-world networks is provided in Table S2.
}\label{fig:realNet}
\end{figure*}

Another unexpected feature that distinguishes the percolation process formulated here from other percolation-like phenomena is the cluster size distribution near criticality. Typically, for second-order phase transitions, in the vicinity of the phase transition point, the size distribution of small connected clusters is in general governed by the monotonous function of $ P(s) \sim s^{-\tau}  e^{-s/ s^*} $, where $ s^* $ provides a characteristic size of the finite components \cite{Newman.Strogatz.Watts-PRE-2001, hu2018local}. The closer to the critical point, the larger $ s^* $ will be. At the exact phase transition point, $ s^* $ approaches infinity and $ P (s) $ exhibits a monotonic power law distribution of $ P (s) \sim s^{-\tau} $, signifying a loss of characteristic scale in the distribution. However, for induced percolation on undirected networks, we find that near the critical point, $ P (s) $ exhibits a novel oscillatory-like behavior, i.e., it is no longer monotonically decreasing with $s$ (see Fig. \ref{fig_small_cluster_distribution}a, b). 

As it can be seen in the figure, the observed oscillatory-like behavior of $ P(s) $ is more pronounced for small values of $s$ and does not change the asymptotic power law distribution for large $s$ nor the critical exponent of the phase transition, which is the same as in bond percolation, $ \beta = 1 $, $ \tau = 5/2 $ \cite{costa2010explosive}. This behavior of $ P(s) $ is, however, clearly distinct from the classical monotonic distribution (see Fig. \ref{fig_small_cluster_distribution}c, d). We note that we don't have a clear notion of what is the exact impact of this pattern on the macroscopic behavior of the system, which is a question to be further examined in future works.

\section{Conclusion}
Motivated by empirical evidences that point to the existence of indirect influence in ecological and social networks, we have proposed a novel percolation model, referred to as induced percolation. This mechanistic process enables a mechanism through which influence propagates beyond nearest neighbors. We found that such indirect interactions lead to a plethora of percolation transitions in complex networks that are rooted in the degree of anisotropy of the connectivity pattern. Specifically, we have shown that the amount of directed links in a network determines the order of the phase transition, which spans from a second order in networks without directed links, to first order when all links are directed. In between, a rich behavior associated to a hybrid phase transitions emerges with the coexistence of second- and first-order phase transitions. Our results imply that the indirect influence between neighbors - by leveraging interaction asymmetry -  fundamentally changes the nature of the phase transition of the system.  In addition, we have reported that the indirect effect makes the size distribution of small clusters near the phase transition point exhibit a non-monotonic pattern, which has not been previously seen in other percolation models. 

On work also represents a step ahead in the theoretical characterization of percolation-like phenomena and in the analytical description that will allow to study the impact of asymmetry of interactions and higher-order correlations on the dynamics of complex networked systems. Direct implications of our model include the fact that first-order transitions could be more abundant than expected in natural and human-made systems where directionality is found more often than not. Finally, we believe that our model represents the first step towards modeling and understanding in more detail the role of interactions that go beyond direct neighbors. Our theoretical framework provides the tools to explore in more depth different scenarios that could realistically describe different types of indirect spreading/influencing mechanisms and their associated macroscopic dynamical behavior.


\begin{table*}[ht]\centering\label{tab:comparison_percolation_models}
\caption{Comparison of percolation models. ``-" indicates that no related research has been found.}
\begin{tabular}{cccccc}\Xhline{2\arrayrulewidth}
	Percolation  & \multicolumn{2}{c}{Type of phase transition} & Clusters distribution & \multirow{2}{*}{$ \beta $} & Hybrid phase transition \\  
	model & undirected & directed &near critical point&&at critical point\\		
	\hline
	
		Induced percolation&2nd &1st &Non-monotonic&\makecell[c]{$ 1 $ (2nd)\\$ 1/2 $ (1st)} & \makecell[c]{$ \theta =1/3 $\\$ \eta= 1/2 $} \\
		
	Bond percolation\cite{callaway2000network,Dorogovtsev2008critical,castellano2009statistical,Cohen.Havlin-2010}&2nd &2nd &monotonic&1&-\\
	
	Site percolation\cite{callaway2000network,albert2000error,Dorogovtsev2008critical,castellano2009statistical,Cohen.Havlin-2010}&2nd &2nd &monotonic&1&-\\ 
	
	Bootstrap percolation\cite{baxter2010bootstrap}&2nd/1st &-&monotonic&\makecell[c]{$ 1 $ (2nd) \\$ 1/2 $ (1st)}&\makecell[c]{$ \theta =1/3 $\\$ \eta= 1/2 $}\\ 
	
	k-core percolation\cite{Dorogovtsev.etal-PRL-2006}&2nd/1st &2nd/1st &-&\makecell[c]{$ 1 $ (2nd) \\$ 1/2 $ (1st)}&-\\ 
	Core percolation\cite{Bauer.Golinelli-EPJB-2001,Liu.Csoka.Zhou.Posfai-PRL-2012}&2nd  &2nd/1st  &\multirow{2}{*}{-}&\makecell[c]{$ 1 $ (2nd) \\$ 1/2 $ (1st)}&-\\

	Explosive percolation\cite{costa2010explosive, riordan2011explosive, grassberger2011explosive}&2nd&-&-&$ 0.0555 $&-\\ 
	
	Articulation percolation\cite{tian2017articulation}&2nd/1st&-&-&\makecell[c]{$ 1 $ (2nd)\\$ 1/2 $ (1st)}&-\\
	
	 \Xhline{2\arrayrulewidth}
\end{tabular}
\end{table*}

\cmttt{\section*{Methods}}
\textbf{Induced percolation on undirected networks.} We elaborate on the definition and the theoretical derivation of induced percolation on undirected networks. All nodes in an undirected network are initially set to state $ 1 $. A node $ l $ remains in state $ 1 $ if at least one of its undirected links has a node $ j $ in state $ 1 $, and this node $ j $ has at least $ m $ neighbors (excluding the node $ l $) with state $ 1 $ (as illustrated in Figure S1A for the case of $ m=2 $); otherwise node $ l $ changes to state $ 0 $ at the next time step. To theoretically analyze the percolating probability that any node belongs to GCC (equivalent to GOUT), $ P_\infty $, we start by defining $ 6 $ conditional probabilities as intermediate variables, whose notations are shown collectively in Table S1 and Figure S2. Without loss of generality, we denote a randomly chosen undirected link as $\left\{j,l\right\}$ and deduce the probability that node $ l $ belongs to GCC.

According to the definition of induced percolation for undirected networks, the condition for node $l$ to remain active (in state $ 1 $) is that there is at least one active neighbor $j$, and the number $\tilde{u}$ of active neighbors (except node $ l $) of node $ j $ satisfies $\tilde{u} \geq m$. We refer to a node in state $ 1 $ as an active node and in state $ 0 $ as an inactive node. Unlike active neighbors in directed networks, the number $\tilde{u}$ of active neighbors in undirected networks is closely related to the degree $ k $ of node $ j $. Specifically, if $k>m$ and node $j$ is active, then node $ j $ can keep all its neighbors active. Conversely, if $k \leq m$, then node $ j $ can not keep any of its neighbors active. Hereafter, we employ the degree $ k $ instead of the number of active neighbors $ \tilde{u} $ to derive percolation probability.

The conditional probability $ \tilde{v} $ is the probability that node $ j $ can keep node $ l $ active (in state 1), given node $ l $ can keep $ j $ active. As per the definition of induced percolation, the event of $ j $ keeping $ l $ active implies that the degree of node $ j $ satisfies $ k > m $. Node $ j $ simultaneously keeps all of its neighbors active. The above analysis yields the following recursive equation:
\begin{equation}\label{eq:undi_v}
\tilde{v} = \sum_{k=m+1}^{\infty} \frac{kP(k)}{\left<k\right>}.
\end{equation}
where $ \frac{kP(k)}{\left<k\right>} $ represents the excess degree distribution of the end node of a randomly chosen link.

On the other hand, the conditional probability $\tilde{v}_{\infty}$ is defined as the probability that node $ j $ can keep node $ l $ active (in state 1), and node $ l $ is connected to the GCC via node $ j $, given that node $ l $ can keep $ j $ active. Again, as per the definition of induced percolation, the degree of node $ j $ satisfies $ k> m $. Analogously, node $ j $ can keep all its neighbors active. In addition, the event that node $ l $ connects to GCC through node $ j $ is equivalent to the event that node $ j $ connects to GCC through at least one of the $ k-1 $ neighbors other than $ l $. The corresponding probability is $ 1 - (1-\tilde{t}_{\infty}-\tilde{v}_{\infty})^{k-1}$ (as shown in Figure S2c), where the probability $ 1-\tilde{t}_{\infty}-\tilde{v}_{\infty} $ accounts for the likelihood that one of the $ k-1 $ neighbors does not belong to the GCC given that node $ j $ can keep it active. Therefore, the self-consistent equation for the conditional probability $\tilde{v}_{\infty}$ can be written as
\begin{equation}\label{eq:undi_v_inf}
\tilde{v}_{\infty} = \sum_{k=m+1}^{\infty} \frac{kP(k)}{\left<k\right>} \left[ 1 - (1-\tilde{t}_{\infty}-\tilde{v}_{\infty})^{k-1} \right].
\end{equation}

In the previous definition, we made use of the conditional probability $\tilde{t}_{\infty}$, which is the probability that node $ j $ cannot keep node $ l $ active (in state 1) while node $ l $ connects to GCC through node $ j $, under the condition that node $ l $ maintains node $ j $ in state 1. Thus, it follows that the degree of node $ j $ satisfies $ k \leq m $ and that node $ j $ cannot keep any of its neighbors active. Moreover, the event in which node $l$ connects to the GCC through node $j$ is equivalent to the event in which node $j$ reaches the GCC through one of the $ k-1 $ neighbors other than $l$. The corresponding probability reads $1-(1-\tilde{a}_{\infty}-\tilde{y}_{\infty})^{k-1 }$ (as shown in Figure S2d), where the probabilities $ \tilde{a}_{\infty}$, $\tilde{y}_{\infty} $ stand for cases in which node $j$ cannot keep any neighbors in state $ 1 $, see below. Therefore, the conditional probability $\tilde{t}_{\infty}$ can be calculated using
\begin{equation}\label{eq:undi_t}
\tilde{t}_{\infty} = \sum_{k=1}^m \frac{kP(k)}{\left<k\right>} \left[1 - (1-\tilde{a}_{\infty}-\tilde{y}_{\infty})^{k-1} \right].
\end{equation}

Once the above probabilities have been defined, we can proceed with the derivation of the remaining three conditional probabilities, namely, $ \tilde{y} $, $\tilde{a}_{\infty}$, $\tilde{y}_{\infty}$, which are analogous to $v $, $ \tilde{v}_{\infty} $ and $\tilde{t}_{\infty} $, but under the condition that node $l $ cannot keep node $j$ active (in state 1). The derivation of the probability $\tilde{y}$ is similar to $\tilde{v}$, except that node $j $ relies on at least one of $k-1 $ neighbors (except $l$) to remain active. This probability can be expressed as
\begin{equation}\label{eq:undi_y}
\tilde{y} = \sum_{k=m+1}^{\infty} \frac{kP(k)}{\left<k\right>} \left[ 1 - (1-\tilde{v})^{k-1} \right].
\end{equation}

The derivation of the conditional probability $\tilde{a}_{\infty}$ is similar to that of $\tilde{t}_{\infty}$, except that one additional condition is required: of the $ k-1 $ neighbors different to $ l $, at least one can keep $j$ active and connected to the GCC. Assuming that there are exactly $ s $ ($ 1 \leq s \leq k-1 $) neighbors that can keep node $ j $ active, the probability is $ \binom{k-1}{s} \tilde{y}^s (1-\tilde{y})^{k-1-s} $. The probability that node $ j $ is connected to GCC through one of the $ s $ neighbors is $\frac{\tilde{y}_{\infty}}{\tilde{y}}$. For the remaining $k-1-s$ neighbors that cannot keep node $j$ active, the probability of $j$ connecting to GCC through one of them is $\frac{\tilde{a}_{\infty}}{ 1-\tilde{y}}$. Therefore, the probability that node $j$ is not connected to GCC through any neighbor is $(1-\frac{\tilde{y}_{\infty}}{\tilde{y }})^s (1-\frac{\tilde{a}_{\infty}}{1-\tilde{y}})^{k-1-s} $, as shown in Figure S2f. Therefore, the self-consistent equation to derive the conditional probability $\tilde{a}_{\infty}$ is
\begin{equation}\label{eq:undi_a}
\begin{aligned}
\tilde{a}_{\infty} =& \sum_{k=1}^{m} \frac{kP(k)}{\left<k\right>} \sum_{s=1}^{k-1} \binom{k-1}{s} \tilde{y}^s (1-\tilde{y})^{k-1-s} \\
& \times \left[1 - (1-\frac{\tilde{y}_{\infty}}{\tilde{y}})^s (1-\frac{\tilde{a}_{\infty}}{1-\tilde{y}})^{k-1-s} \right] \\
=& \sum_{k=1}^{m} \frac{kP(k)}{\left<k\right>}\left[1 - (1-\tilde{y})^{k-1} - (1-\tilde{y}_{\infty}-\tilde{a}_{\infty})^{k-1} \right. \\
&\left.+ (1-\tilde{y}-\tilde{a}_{\infty})^{k-1} \right].
\end{aligned}
\end{equation}

Finally, the conditional probability $\tilde{y}_{\infty}$ can be obtained similarly to $\tilde{v}_{\infty}$, with the additional consideration that for $ k-1 $ neighbors except $ l $, at least one can keep $j$ active and that node $ j $ connects to the GCC via at least one of the $ k-1 $ neighbors. Thus, the degree of node $ j $ satisfies $ k> m $, which also implies that $ j $ keeps all its neighbors active. Therefore, the conditional probabilities $ \tilde{y} $, $ \tilde{y}_{\infty} $, $ \tilde {a}_{\infty} $ in Eq.~(\ref{eq:undi_a}) are replaced by probabilities $ \tilde{v} $, $ \tilde{v}_{\infty} $, $ \tilde{t}_{ \infty} $. This leads to the following expression for the conditional probability $\tilde{y}_{\infty}$ 
\begin{equation}\label{eq:undi_y_inf}
\begin{aligned}
\tilde{y}_{\infty} =& \sum_{k=m+1}^{\infty} \frac{kP(k)}{\left<k\right>}  \sum_{s=1}^{k-1} \binom{k-1}{s} \tilde{v}^s (1-\tilde{v})^{k-1-s} \\
&\left[1 - (1-\frac{\tilde{v}_{\infty}}{\tilde{v}})^s (1-\frac{\tilde{t}_{\infty}}{1-\tilde{v}})^{k-1-s} \right] \\
=& \sum_{k=m+1}^{\infty} \frac{kP(k)}{\left<k\right>}  \left[ 1 - (1-\tilde{v})^{k-1} - (1-\tilde{t}_{\infty}-\tilde{v}_{\infty})^{k-1} \right.\\
&\left.+ (1-\tilde{v}-\tilde{t}_{\infty})^{k-1} \right],
\end{aligned}
\end{equation}
where $ s $ represents the number of neighbors that can keep $ j$ active. The graphical solution of the self-consistent equation $\tilde{y}_{\infty}$ is shown in the main text, where $f(\tilde{y}_{\infty}) = F(\tilde{y}_{\infty} )-\tilde{y}_{\infty}$ and $F(\tilde{y}_{\infty})$ represents the expression on the right hand side of equation Eq.(\ref{eq:undi_y_inf}). The value of $F(\tilde{y}_{\infty})$ is obtained by solving the self-consistent equations Eq.(\ref{eq:undi_v})-Eq.(\ref{eq:undi_a}).

The previously defined conditional probabilities allow to derive the order parameter, $P_{\infty}$, for induced percolation on undirected networks. For an arbitrarily chosen node $ l $ to belong to the GCC, we have that i) at least one of its neighbors should keep it active, and ii) node $ l $ is attached to the GCC through at least one of its neighbors. If the degree of node $ l $ satisfies $ k \leq m $, then the probability that node $ l $ belongs to GCC is $1-(1-\tilde{y})^k-(1-\tilde{y}_{\infty}-\tilde{a}_{\infty})^k + (1-\tilde{y}-\tilde{a}_{\infty})^k$, whose derivation is similar to Eq.~(\ref{eq:undi_a}) in $\tilde{a}_{\infty}$. If the degree of node $ l $ satisfies $ k> m $, then the probability that node $ l $ belongs to the GCC is $1-(1-\tilde{v})^k-(1-\tilde{t}_{\infty}-\tilde {v}_{\infty})^k + (1-\tilde{v}-\tilde{t}_{\infty})^k$ and the derivation is similar to $\tilde{y}_{\infty}$ in Eq.~(\ref{eq:undi_y_inf}). Therefore, the order parameter $P_{\infty}$ can be computed, for undirected networks, as
\begin{widetext}
\begin{equation}\label{eq:undi_giant}
\begin{aligned}
P_{\infty} &= \sum_{k=0}^m P(k) \left[1 - (1-\tilde{y})^k - (1-\tilde{y}_{\infty}-\tilde{a}_{\infty})^k + (1-\tilde{y}-\tilde{a}_{\infty})^k \right] \\
&+ \sum_{k=m+1}^{\infty} P(k) \left[1 - (1-\tilde{v})^k - (1-\tilde{t}_{\infty}-\tilde{v}_{\infty})^k + (1-\tilde{v}-\tilde{t}_{\infty})^k \right]. \\
\end{aligned}
\end{equation}
\end{widetext}

\section{Supplementary Information: Induced Percolation on Networked Systems}

\subsection{Induced percolation on mixed networks}
\begin{figure}[!htp]
	\centering
	\includegraphics[width=.7\textwidth]{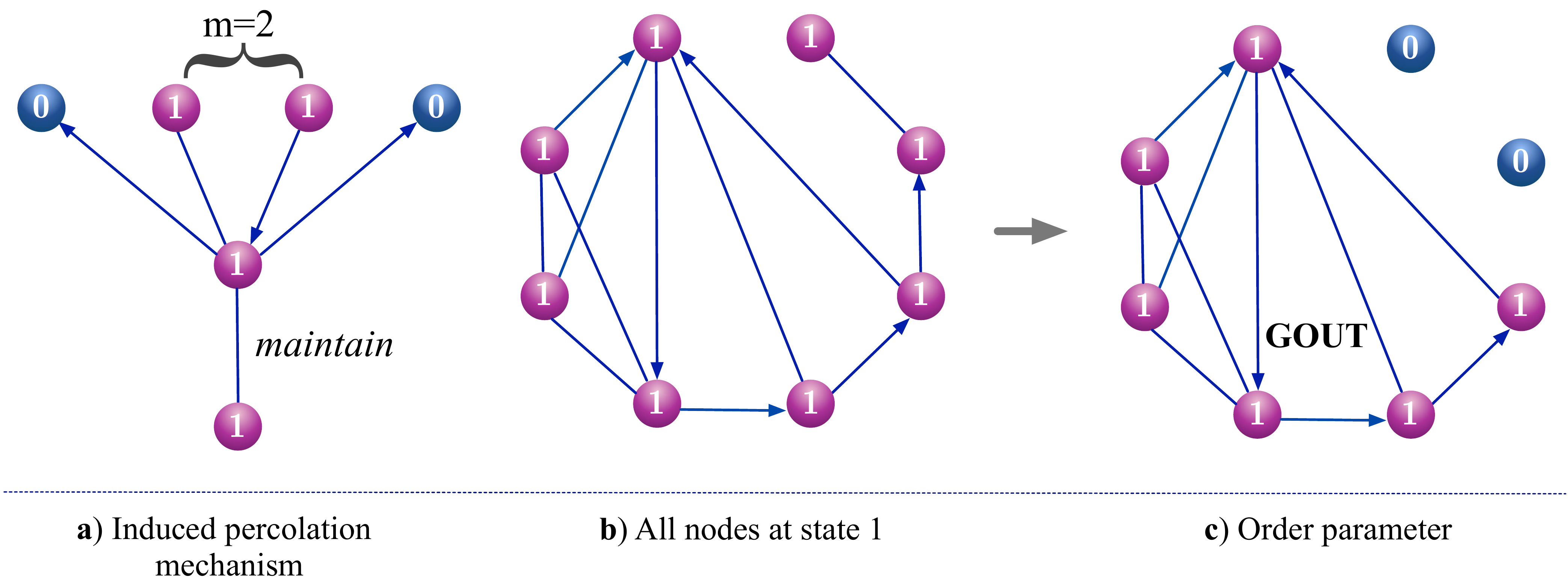}%
	\caption{Induced percolation on mixed networks.}%
	\label{fig_inducing_mixed}%
\end{figure}

Induced percolation on mixed networks with a mix of directed and undirected links is defined as follows. All nodes are initially set to state $ 1 $. A node $ l $ remains in state $ 1 $ if at least one of its undirected or directed links has a node $ j $ in state $ 1 $, and this node $ j $ has at least $ m $ neighbors \cmt{(including in-coming neighbors and undirected neighbors while excluding node $ l $)} with state $ 1 $ (as illustrated in Figure \ref{fig_inducing_mixed}a for the case of $ m=2 $); otherwise node $ l $ changes to state $ 0 $ at the next time step. The formulation of the induced percolation framework on mixed networks is more complicated than that on directed and undirected networks. One of the main challenges is given by the intertwined effect of directed and undirected edges in maintaining neighbors state and connecting neighbors to the GOUT, which significantly increases the possibilities for a node connecting to GOUT. For example, a mixed network needs to consider staying active through directed neighbors and connecting to GOUT through undirected neighbors, or staying active through undirected neighbors and connecting to GOUT through directed outgoing links, which is not the case for directed and undirected networks. Therefore, in order to deal with this intertwined issue, we employ a multilayer network approach to separate the directed and undirected neighbors by layers. Thus, nodes on the same layer maintain the same pattern in relation to them being active, which enables a recursive calculation. Through aggregation of network layers, we account for all the possibilities in the calculation of the order parameter while effectively avoiding the previous challenge of dealing concurrently with directed and undirected links.

To derive the order parameter for induced percolation on mixed networks, we thus represent a mixed network by a multiplex network: one layer includes undirected links and the other layer includes directed links. A mixed random network is generated by assigning a single direction to an undirected link chosen from an undirected random network. The proportion $ p $ of assigned directed links and the average degree $ \left<k\right> $ of the underlying undirected network constitute the parameter space for the generation of mixed networks. In this setting, the average degree in the undirected layer $ \left<k\right>_u $ and the average degree in the directed layer $ \left<k\right>_d $ follow $ \left<k\right>_u = \left<k\right>(1-p) $ and $ \left<k\right>_d = \frac{ \left<k\right> p} {2} $. By randomly selecting a directed link in the directed layer, the excess degree distribution of the starting node can be written as $\frac{k_o P(k,k_i,k_o)}{ \left<k\right>_d }$, where $ k_i$, $k_o $ represent the in-degree and out-degree in the directed layer, respectively, and $ k $ is the degree in the undirected layer. Similarly, if an undirected link is randomly chosen in the undirected layer, the excess degree distribution of its end node follows $ \frac{k P(k,k_i,k_o)}{ \left<k\right>_u }$. Following the same arguments as for the case of undirected networks, we use the undirected degree $ k $ of a node $ l $ to deduce the percolation probability, instead of using the number of undirected active neighbors.

On mixed networks, we mainly consider GOUT as the order parameter. For an active node $l$ to be in GOUT, at least one active in-coming or undirected neighbor should belong to the GOUT. For a randomly chosen node $ l $ to remain in state $ 1 $, the definition of induced percolation implies that there is at least one active neighbor $j$ in the directed layer or at least one active neighbor $j$ in the undirected layer, and node $j$ has $ s $ active incoming neighbors in the directed layer and $ \tilde{u} $ active neighbors in the undirected layer (excluding node $ l $) satisfying $s + \tilde{u} \geq m$. As noted, given that the number $\tilde{u}$ of active neighbors is closely associated with the undirected degree $ k $ of node $ j $, we use $ k $ instead of $ \tilde{u} $ in deriving GOUT. 

To obtain the condition for a node $ j $ to keep its neighbors active, we consider three cases separately: (i) If $ s + k> m $, then $ j $ can maintain neighbors either in the directed layer or in the undirected layer active; (ii) If $ s + k <m $, then $ j $ cannot keep any neighbors active; (iii) if $ s + k = m $, then node $ j $ cannot keep its undirected neighbors active, but node $ j $ can keep its outgoing neighbors in the directed layer active under the additional condition that all undirected neighbors of $ j $ are active ($ \tilde{u} =k $).  

We start with the definition of a list of intermediate conditional probabilities, as described in Table \ref{tab:mixed}. The order parameter $ P_\infty $ is eventually obtained based on solutions of the defined probabilities. We present the definition and the solution of intermediate conditional probabilities one at a time. 
\begin{table}
	\centering
	\caption{\cmt{Definition of conditional probabilities used to derive the probability of induced percolation on mixed networks. Each row shows one conditional probability. The first column shows the notation used to denote the conditional probability that the event in the fourth column occurs for a randomly selected undirected link $ \{j,l \}$ or a directed link $ (j,l)$, given the conditional event presented in the third column. The right arrow ($ j \rightarrow l$) represents that node $ j $ can keep node $ l $ active, otherwise denoted as $j \nrightarrow l $. For example, the conditional probability $\tilde{v}_{\infty}$ shows the probability that given node $ l $ can keep node $ j $ active: (i) node $ j $ can keep node $ l $ active and (ii) $ l $ connects to GOUT through node $ j $.}}
	\label{tab:mixed}
	\begin{tabular}{c|c|c|c|c|c}
		\hline
		Conditional probability & Link type & Conditional event & \multicolumn{2}{c|}{Occuring event} & Relation \\\hline\hline
		
		$y$ & \multirow{3}{*}{\makecell[c]{ Directed\\$(j,l)$}} & \multirow{3}{*}{None} & \multicolumn{2}{c|}{$j \rightarrow l$} & \\ \cline{1-1}\cline{4-6}
		
		$a$ & & & \multicolumn{2}{c|}{$j$ is active but $j \nrightarrow l$} & \\ \cline{1-1}\cline{4-6}
		$x$ & & & \multicolumn{2}{c|}{$j$ is active} & $x=y+a$ \\ \hline
		
		$\tilde{v}$ & \multirow{4}{*}{\makecell[c]{Undirected \\ $ \{j,l\}$}} & $l \rightarrow j$ & \multicolumn{2}{c|}{$j \rightarrow l$} & \\ \cline{1-1} \cline{3-6}
		$\tilde{y}$ &  & $l \nrightarrow j$& \multicolumn{2}{c|}{$j \rightarrow l$} & \\ \cline{1-1} \cline{3-6}
		$\tilde{a}$ &  & $l \nrightarrow j$ & \multicolumn{2}{c|}{$j$ is active but $j \nrightarrow l$} & \\ \cline{1-1} \cline{3-6}
		$\tilde{x}$ &  & $l \nrightarrow j$ & \multicolumn{2}{c|}{$j$ is active} & $\tilde{x} = \tilde{y}+\tilde{a}$ \\ \hline

		$y_{\infty}$ & \multirow{3}{*}{\makecell[c]{Directed\\$(j,l)$}} & \multirow{3}{*}{None} & (i) $j \rightarrow l$ & \multirow{8}{*}{\makecell[c]{and (ii) $l$ \\connects to \\ GOUT via $j$}} &\\\cline{1-1}\cline{4-4}\cline{6-6}
		$a_{\infty}$ & & & (i) $j \nrightarrow l$ & & \\\cline{1-1}\cline{4-4}\cline{6-6}
		$x_{\infty}$ & & & (i) $j$ is active & & $x_{\infty} = y_{\infty} + a_{\infty}$\\\cline{1-4}\cline{6-6}
		
		$\tilde{t}_{\infty}$ & \multirow{5}{*}{\makecell[c]{Undirected \\ $ \{j,l\}$}} & $l \rightarrow j$ & (i) $j \nrightarrow l$ & &\\\cline{1-1}\cline{3-4}\cline{6-6}
		$\tilde{v}_{\infty}$ &  & $l \rightarrow j$ & (i) $j \rightarrow l$ &  &\\\cline{1-1}\cline{3-4}\cline{6-6}
		$\tilde{y}_{\infty}$ &  & $l \nrightarrow j$ & (i) $j \rightarrow l$ &&\\\cline{1-1}\cline{3-4}\cline{6-6}
		$\tilde{a}_{\infty}$ &  & $l \nrightarrow j$ & (i) $j \nrightarrow l$ &&\\\cline{1-1}\cline{3-4}\cline{6-6}
		$\tilde{x}_{\infty}$ &  & $l \nrightarrow j$ & (i) $j$ is active && $\tilde{x}_{\infty} = \tilde{y}_{\infty} + \tilde{a}_{\infty}$\\\hline
	\end{tabular}
\end{table}

The first probability $ y $ is defined as the probability that node $ j $ can keep node $ l $ active through a randomly selected directed link $ (j,l) $. As per the definition of induced percolation, node $ j $ can keep node $ l $ active if $ s + k \geq m $ and $j$ is active. In the case of $ s+k > m $, node $ j $ is kept active by at least one of the $s$ active in-coming neighbors or $k$ undirected neighbors with probability $1-\left(\frac{ a} {x}\right)^s(1- \tilde{v})^k $. In the case of $ s + k = m $, only when all the undirected neighbors are active, with probability $ \tilde{x}^k $, node $ j $ can keep node $ l $ active. From the above analysis, the equation for solving the probability $ y $ reads
\begin{equation}\label{eq:mixed_y}
\begin{aligned}
y &= \sum_{k, k_i, k_o}^{\infty} \frac{k_o P(k,k_i,k_o)}{ \left<k\right>_d } \sum_{s=0}^{k_i} \binom{k_i}{s} (1-x)^{k_i-s} x^s  \\
&\times\Bigg\{ \mathbb{I}(s + k>m) \left[1 - \left( \frac{a}{x} \right)^s (1-\tilde{v})^k \right] +
\mathbb{I}(s + k=m)  \left[ \tilde{x} ^k - \tilde{a}^k\left( \frac{a}{x} \right)^s \right] \Bigg\} 
\end{aligned}
\end{equation}
where $ \mathbb{I}(x)$ represents the indicator function of the logical statement $ x $: $ \mathbb{I}(x)=1 $ if $ x $ is true, otherwise $ \mathbb{I}(x) = 0 $.

For the probability $ a $, the degree of node $ j $ satisfies $ s + k \leq m $. When $ s+k<m $, node $ j $ cannot keep node $l$ active. The probability $ a $ thus reduces to the probability of node $ j $ remaining active under the condition of having $ s $ active in-coming neighbors, which is $1-\left( \frac{a}{x} \right)^s (1-\tilde{y})^k$. When $ s+k=m $, \cmt{the probability that node $ j $ can not keep node $ l $ active} follows $1-\left( \frac{a}{x} \right)^s (1-\tilde{y})^k-\tilde{x} ^k + {\tilde{a}} ^k \left( \frac{a}{x} \right)^s$, where $\left( \frac{a}{x} \right)^s (1-\tilde{y})^k$ is the probability of $j$ being \cmt{inactive (in state 0)}, and $\tilde{x} ^k$ is the probability of all undirected neighbors being active. The term $\left( \frac{a}{x} \right)^s {\tilde{a}} ^k$ gives the probability that node $ j $ is \cmt{inactive} and at the same time all undirected neighbors are active too. According to the above analysis, the probability $ a $ can be obtained as	
\begin{equation}\label{eq:mixed_a}
\begin{aligned}
a &= \sum_{k, k_i, k_o}^{\infty} \frac{k_o P(k,k_i,k_o)} {\langle k \rangle _d} \sum_{s=0}^{k_i} \binom{k_i}{s} (1-x)^{k_i-s} x^s \\
& \times \left\{ \mathbb{I}(s+k \leq m) \left[1 - \left( \frac{a}{x} \right)^s (1-\tilde{y})^k \right] - \mathbb{I}(s+k = m) \left[ \tilde{x} ^k- {\tilde{a}} ^k \left( \frac{a}{x} \right)^s    \right] \right\}.
\end{aligned}
\end{equation}

For the conditional probability $\tilde{v}$, the node $ j $ can keep node $ l $ active if $ s + k> m $ and $j$ is active, which also indicates that $ j $ can keep all undirected neighbors active. Therefore, the equation for solving the probability $\tilde{v}$ is
\begin{equation}\label{eq:mixed_tildeV}
\tilde{v} = \sum_{k, k_i, k_o}^{\infty} \frac{k P(k,k_i,k_o)}{ \left<k\right>_u } \sum_{s=0}^{k_i} \mathbb{I}(s + k>m)\binom{k_i}{s} (1-x)^{k_i-s} x^s 
\end{equation}

The derivation of the conditional probability $\tilde{y}$ is similar to $\tilde{v} $, except that node $ l $ cannot keep node $j$ active. The additional requirement in calculating probability $\tilde{y}$ is that node $ j $ is kept active by at least one of $s$ active neighbors in the directed layer and $k-1$ neighbors except $l$ in the undirected layer. The corresponding probability follows $1-\left( \frac{a}{x} \right)^s(1-\tilde{v})^{k-1}$. Therefore, the corresponding equation to solve $\tilde{y} $ is
\begin{equation}
\tilde{y} = \sum_{k, k_i, k_o}^{\infty} \frac{k P(k,k_i,k_o)}{ \left<k\right>_u } \sum_{s=0}^{k_i} \mathbb{I}(s + k>m) \binom{k_i}{s} (1-x)^{k_i-s} x^s \left[ 1-\left(\frac{a}{x}\right)^{s}(1-\tilde{v})^{k-1} \right] 
\end{equation}

For the conditional probability $ \tilde{a} $, the event that node $ j $ is active but cannot keep node $ l $ active indicates $ s + k \leq m $. Therefore, the solution to the probability $ \tilde{a} $ is similar to Eq.~(\ref{eq:mixed_a}):
\begin{equation}
\tilde{a} = \sum_{k, k_i k_o}^{\infty} \frac{k P(k,k_i,k_o)}{ \left<k\right>_u } \sum_{s=0}^{k_i} \mathbb{I}(s + k \leq m)\binom{k_i}{s} (1-x)^{k_i-s} x^s\left[ 1-\left(\frac{a}{x}\right)^{s}(1-\tilde{y})^{k-1} \right] 
\end{equation}

The derivation of the probability $y_{\infty}$ is analogous to the derivation of the probability $ y $, while an additional requirement is that at least one of the $s$ active neighbors in the directed layer and $k$ neighbors in the undirected layer belong to GOUT. Therefore, the equation to solve $y_{\infty}  $ is expressed as
\begin{equation}\label{eq:mixed_y_inf}
\begin{aligned}
y_{\infty} &= \sum_{k, k_i, k_o}^{\infty} \frac{k_o P(k,k_i,k_o)}{ \left<k\right>_d } \sum_{s=0}^{k_i}\binom{k_i}{s} (1-x)^{k_i-s} x^s \\
&\times\left[ \mathbb{I}(s + k>m) Y_{1}(s, k)+\mathbb{I}(s + k=m) Y_{2}(s, k)\right] \\
\end{aligned}
\end{equation}
where $Y_1(s,k)$ denotes the probability that $j $ connects to GOUT under the conditions $ s + k> m $ and $j$ being active. In this case, node $j$ can keep all its undirected neighbors active. According to the definition of induced percolation, at least one of $s + k $ neighbors keeps $ j $ active, and $ j $ connects to GOUT through at least one of the $ s + k $ neighbors. Therefore, the probability $Y_1(s,k)$ is calculated as
\begin{equation}\label{eq:mixed_Y1}
\begin{aligned}
Y_{1}(s, k) &=\sum_{d=0}^{s} \sum_{u=0}^{k} \mathbb{I}(d+u>0) \binom{s}{d} \left( \frac{y}{x} \right)^d \left( \frac{a}{x} \right)^{s-d} \binom{k}{u} \tilde{v}^{u}(1-\tilde{v})^{k-u} \\
&\times \left[1 -\left(1-\frac{y_{\infty}}{y}\right)^{d} \left(1-\frac{a_{\infty}}{a}\right)^{s-d} \left(1-\frac{\tilde{v}_{\infty}}{\tilde{v}}\right)^{u} \left(1-\frac{\tilde{t}_{\infty}}{1-\tilde{v}}\right)^{k-u}\right] \\
&=1- \left( 1 -\frac{x_{\infty}}{x} \right)^{s} (1-\tilde{v}_{\infty}-\tilde{t}_{\infty})^{k} 
- \left( \frac{a}{x} \right)^{s}(1-\tilde{v})^{k} + \left( \frac{a-a_{\infty}}{x} \right)^{s} (1-\tilde{v}-\tilde{t}_{\infty})^{k}
\end{aligned}
\end{equation}

In addition, $Y_2(s,k)$ represents the probability that (i) node $ j $ can keep its out-going neighbor $ l $ active and (ii) $ j $ connects to GOUT, given the condition $ s + k = m $. In this case, node $ j $ can keep node $l$ active if all undirected neighbors of $ j $ are active. Therefore, $Y_2(s,k)$ is calculated as
\begin{equation}\label{eq:mixed_Y2}
\begin{aligned}
Y_{2}(s, k) &=\sum_{d=0}^{s} \sum_{u=0}^{k} \mathbb{I}(d+u>0) \binom{s}{d} \left( \frac{y}{x} \right)^d \left( \frac{a}{x} \right)^{s-d} \binom{k}{u} \tilde{y}^{u}\tilde{a}^{k-u} \\
&\times \left[1 -\left(1-\frac{y_{\infty}}{y}\right)^{d} \left(1-\frac{a_{\infty}}{a}\right)^{s-d} \left(1-\frac{\tilde{y}_{\infty}}{\tilde{y}}\right)^{u} \left(1-\frac{\tilde{a}_{\infty}}{\tilde{a}}\right)^{k-u}\right] \\
&=\tilde{x}^k - \left( 1 -\frac{x_{\infty}}{x} \right)^{s} (\tilde{x} -\tilde{x}_{\infty})^{k} - \left( \frac{a}{x} \right)^{s}\tilde{a}^{k} + \left( \frac{a-a_{\infty}}{x} \right)^{s} (\tilde{a} -\tilde{a}_{\infty})^{k}
\end{aligned}
\end{equation}

The derivation of the probability $a_{\infty}$ is similar to $y_{\infty}$ and $ a $. The probability $a_{\infty}$ is defined for the event in which $ j $ is active but $ j $ cannot keep $ l $ active, indicating that $ s + k \leq m $. However, in the case of $ s + k = m $ and if all undirected neighbors of node $ j $ are active, node $ j $ can keep node $ l $ active. The corresponding probability, which reads $ \mathbb{I}(s + k=m) Y_{2}(s, k) $, should be subtracted from the probability $ a_{\infty}$. Therefore, the probability $a_{\infty}$ can be written as
\begin{equation}\label{eq:mixed_a_inf}
\begin{aligned}
a_{\infty} &= \sum_{k, k_{i} k_{o}}^{\infty} \frac{k_o P(k,k_i,k_o)} {\langle k \rangle _d} \sum_{s=0}^{k_i} \binom{k_i}{s} (1-x)^{k_{i}-s} x^{s} \\
& \times \left[ \mathbb{I}(s+k \leq m) Y_3(s, k) - \mathbb{I}(s+k=m) Y_2(s,k) \right]. \\
\end{aligned}
\end{equation}
where $Y_3(s,k)$ denotes the probability that node $j $ connects to GOUT through neighbors other than $l$, given that $s + k \leq m $,
\begin{equation}\label{eq:mixed_A}
\begin{aligned}
Y_3(s, k) &=\sum_{d=0}^{s} \sum_{u=0}^{k} \mathbb{I}(d+u>0) \binom{s}{d} \left( \frac{y}{x} \right)^d \left( \frac{a}{x} \right)^{s-d} \binom{k}{u} \tilde{y}^{u}(1-\tilde{y})^{k-u} \\
&\times\left[1- \left(1-\frac{y_{\infty}}{y}\right)^{d} \left(1-\frac{a_{\infty}}{a}\right)^{s-d} \left(1-\frac{\tilde{y}_{\infty}}{\tilde{y}}\right)^{u} \left(1-\frac{\tilde{a}_{\infty}}{1-\tilde{y}}\right)^{k-u}\right] \\
&= 1 - \left( 1 -\frac{x_{\infty}}{x} \right)^{s} (1-\tilde{x}_{\infty})^{k} - \left( \frac{a}{x} \right)^{s} (1-\tilde{y})^{k} + \left( \frac{a-a_{\infty}}{x} \right)^{s}(1-\tilde{y}-\tilde{a}_{\infty})^{k}.
\end{aligned}
\end{equation}

The probability $\tilde{t}_{\infty}$ is defined for the event that node $ j $ cannot keep node $ l $ active and $ j $ connects to GOUT through neighbors other than $ l $, given the condition that $ l $ can keep $ j $ active. The event that $ j $ cannot keep node $ l $ active indicates $ s+k \leq m $. Therefore, we have that
\begin{equation}\label{eq:mixed_tildeT_inf}
\begin{aligned}
\tilde{t}_{\infty} &= \sum_{k, k_i, k_o}^{\infty} \frac{k P(k,k_i,k_o)}{ \left<k\right>_u } \sum_{s=0}^{k_i} \mathbb{I}(s + k \leq m) \binom{k_i}{s} (1-x)^{k_{i}-s} x^{s} \\
& \times \left[1- \left( 1-\frac{x_{\infty}}{x}\right)^{s} \left(1-\tilde{x}_{\infty}\right)^{k-1}\right]
\end{aligned}
\end{equation}

Probabilities $\tilde{v}_{\infty}$, $\tilde{y}_{\infty}$ and $\tilde{a}_{\infty}$ are derived analogously to $\tilde{v}$ , $\tilde{y}$ and $\tilde{a}$, except that additional conditions are required: node $ j $ connects to GOUT through at least one of the $ s $ active in-coming neighbors and $ k-1 $ undirected neighbors (except $ l $). Therefore, the probability $\tilde{v}_{\infty}$ is obtained from
\begin{equation}\label{eq:mixed_tildeV_inf}
\begin{aligned}
\tilde{v}_{\infty} &= \sum_{k, k_i, k_o}^{\infty} \frac{k P(k,k_i,k_o)}{ \left<k\right>_u } \sum_{s=0}^{k_i} \mathbb{I}(s + k>m)\binom{k_i}{s} (1-x)^{k_i-s} x^s \\
&\times \left[ 1 - \left(1 - \frac{x_{\infty}}{x} \right)^s (1 - \tilde{v}_{\infty} - \tilde{t}_{\infty})^{k-1} \right]
\end{aligned}
\end{equation}
and
\begin{equation}\label{eq:mixed_tildeY_inf}
\tilde{y}_{\infty} = \sum_{k, k_{i} k_{o}}^{\infty} \frac{kP(k,k_i,k_o)}{ \left<k\right>_u }\sum_{s=0}^{k_{i}} \mathbb{I}(s + k>m) \binom{k_i}{s} (1-x)^{k_{i}-s} x^{s} Y_1(s, k-1) 
\end{equation}
where $Y_1(s, k-1)$ means that (i) $ j $ can keep the undirected neighbor $ l $ active, and (ii) $ j $ connects to GOUT through nodes other than $ l $, given that $s + k>m$. The value of $Y_1(s, k-1)$ is obtained from the equation Eq.~(\ref{eq:mixed_Y1}).
Analogously, the probability $ \tilde{a}_{\infty} $ is expressed as
\begin{equation}\label{eq:mixed_tildeA_inf}
\tilde{a}_{\infty} = \sum_{k, k_{i} k_{o}}^{\infty} \frac{kP(k,k_i,k_o)}{ \left<k\right>_u } \sum_{s=0}^{k_{i}} \mathbb{I}(s + k \leq m) \binom{k_i}{s} (1-x)^{k_{i}-s} x^{s} Y_3(s, k-1)
\end{equation}
where $Y_3(s, k-1)$ represents the probability that $ j $ connects to GOUT through neighbors other than $ l $, given that $ s + k \leq m $, whose value is determined by the equation Eq.~(\ref{eq:mixed_A}).

Based on the solutions of the above defined probabilities, the order parameter GOUT on mixed networks can be calculated considering two contributions. When the total number of active neighbors $s$ in the directed layer and the number $ k $ of neighbors in the undirected layer satisfies $s+k>m$ and when $j$ is active,  node $j$ can keep all the undirected neighbors active. The probability that $j$ belongs to GOUT is $Y_1(s,k)$. However, when $s+k \leq m$, node $j$ cannot keep any of its undirected neighbors active. The probability that node $j$ belongs to GOUT is $Y_3(s,k)$. Finally, the order parameter GOUT is given by
\begin{equation}
\begin{aligned}
P_{\infty} &= \sum_{k, k_{i} k_{o}}^{\infty} P(k, k_{i}, k_{o}) \sum_{s=0}^{k_{i}} \binom{k_i}{s} (1-x)^{k_{i}-s} x^{s} \\
&\times \left[ \mathbb{I}(s + k>m) Y_{1}(s, k) + \mathbb{I}(s + k \leq m) Y_3(s,k) \right] \\
\end{aligned}
\end{equation}
where $Y_1(s,k)$ and $Y_3(s,k)$ are established in equations Eqs.~(\ref{eq:mixed_Y1}) and (\ref{eq:mixed_A}).

\begin{figure}
	\includegraphics[width=0.6\textwidth]{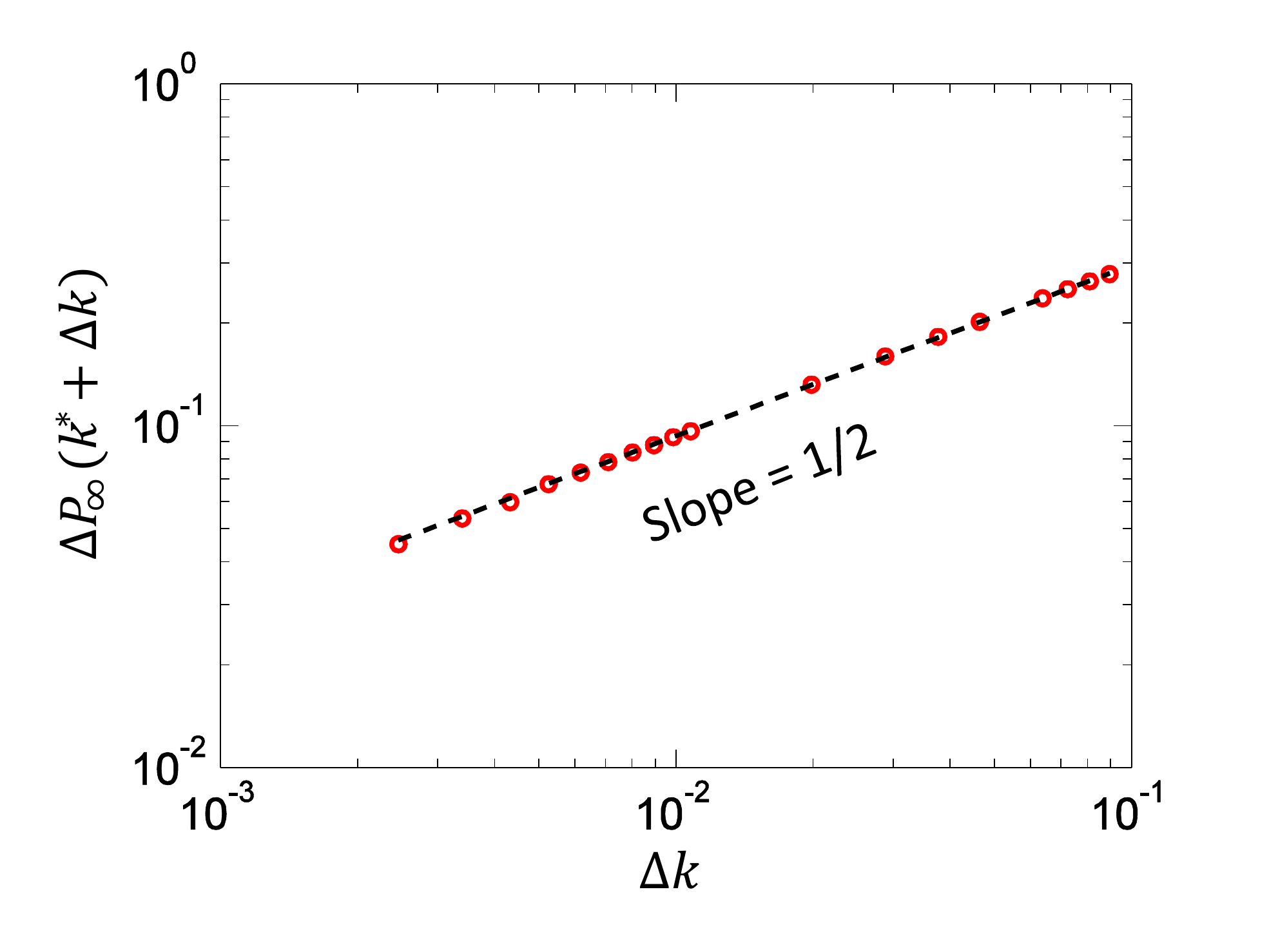}
	\caption{$ \Delta P_\infty $ as a function of $ \Delta k=\left<k\right>-k^* $ when approaching the critical point from above.}
	\label{fig:SI_critical_exponent_beta}
\end{figure}

\begin{figure}
	\includegraphics[width=0.6\textwidth]{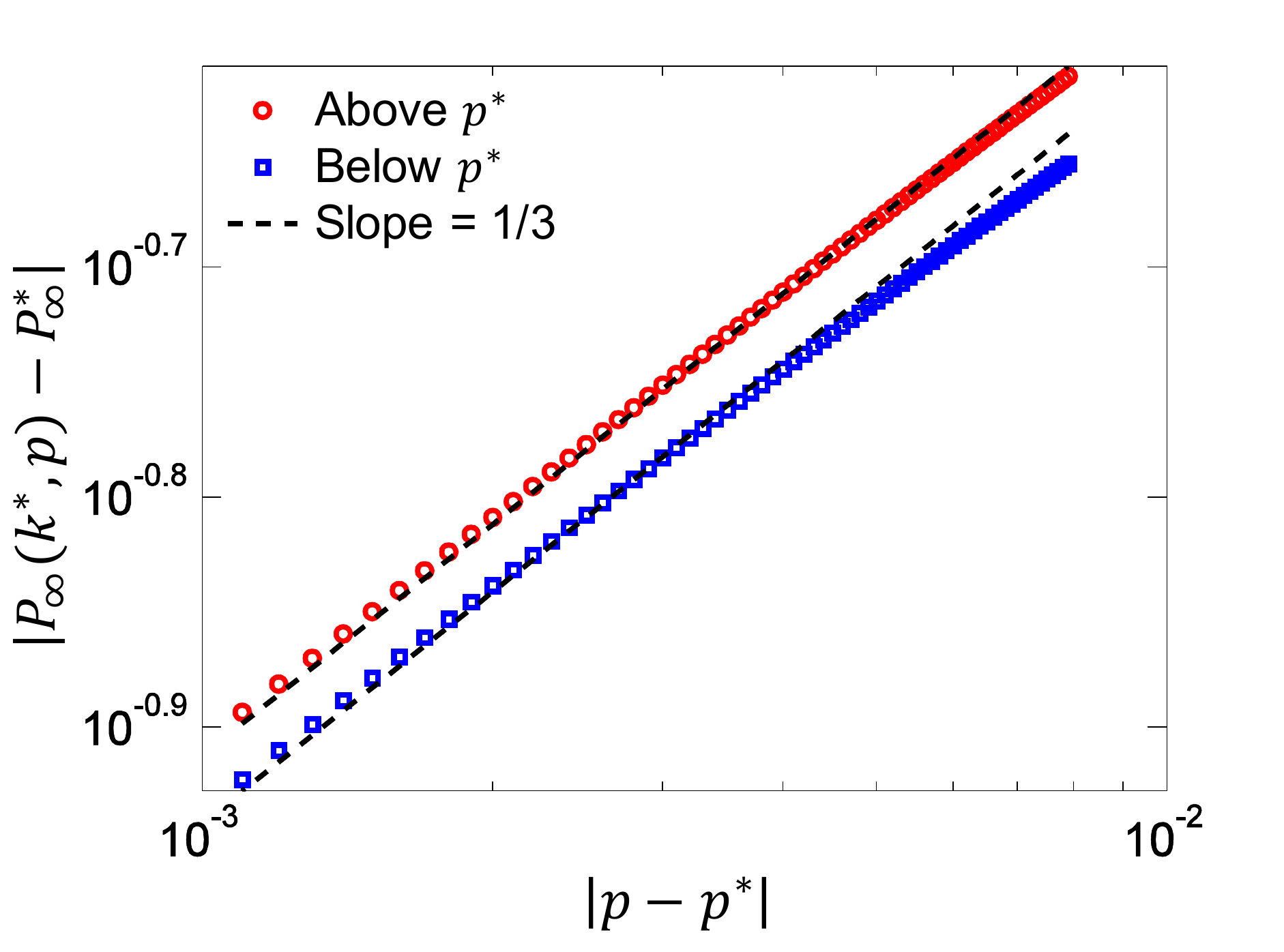}
	\caption{The change of $ P_\infty $ near the critical point $ p^* $ as a function of $ \Delta p=p-p^* $ when fixing $ \left<k\right>=k^* $. }
	\label{fig:SI_critical_exponent_gamma}
\end{figure}

In addition, the order parameter $ P_{\infty} $ on mixed networks exhibits hybrid phase transitions with the presence of certain amount of directed links. Within the hybrid transition, variables of $ P_{\infty} $ follow a set of scaling relations with critical exponents in line with Landau's mean-field theory. Specifically, the size of the jump of GOUT, $\Delta P_\infty \coloneqq \lim_{p \rightarrow p_c^{-}}P_\infty(k^*+\Delta k,p)- \lim_{p \rightarrow p_c^{+}}P_\infty (k^*+\Delta k,p) $, where $ p_c $ is the critical point at which the first order transition occurs, follows a scaling function of $ \Delta k $ with the critical exponent $\eta = 1/2$ (Fig. \ref{fig:SI_critical_exponent_beta})
\begin{equation}
\Delta P_\infty\left(k^*+\Delta k\right)  \sim \left(\Delta k\right)^\frac{1}{2}.
\end{equation}
The scaling relation between $ \Delta P_\infty $ and $ p -p^* $ is presented in the main text. If $ \left<k\right> $ is fixed at $ k^* $ and we vary $ p $ in the vicinity of $ p^* $, the size deviation of GOUT can be quantified by the following scaling function of $ p-p^* $ with critical exponent $ \theta = 1/3 $ (Fig. \ref{fig:SI_critical_exponent_gamma}), reached from both below and above, 
\begin{equation}
\left| P_\infty(k^*,p)- P_\infty^* (k^*,p^*)\right|  \sim \left|p-p^* \right|^{\frac{1}{3}}.
\end{equation}
The scaling behavior of GOUT when fixing $ p $ at $ p^* $ is presented in the main text.

\subsection{Induced percolation on undirected networks}
\begin{figure}[!htp]
	\centering
	\includegraphics[width=.7\textwidth]{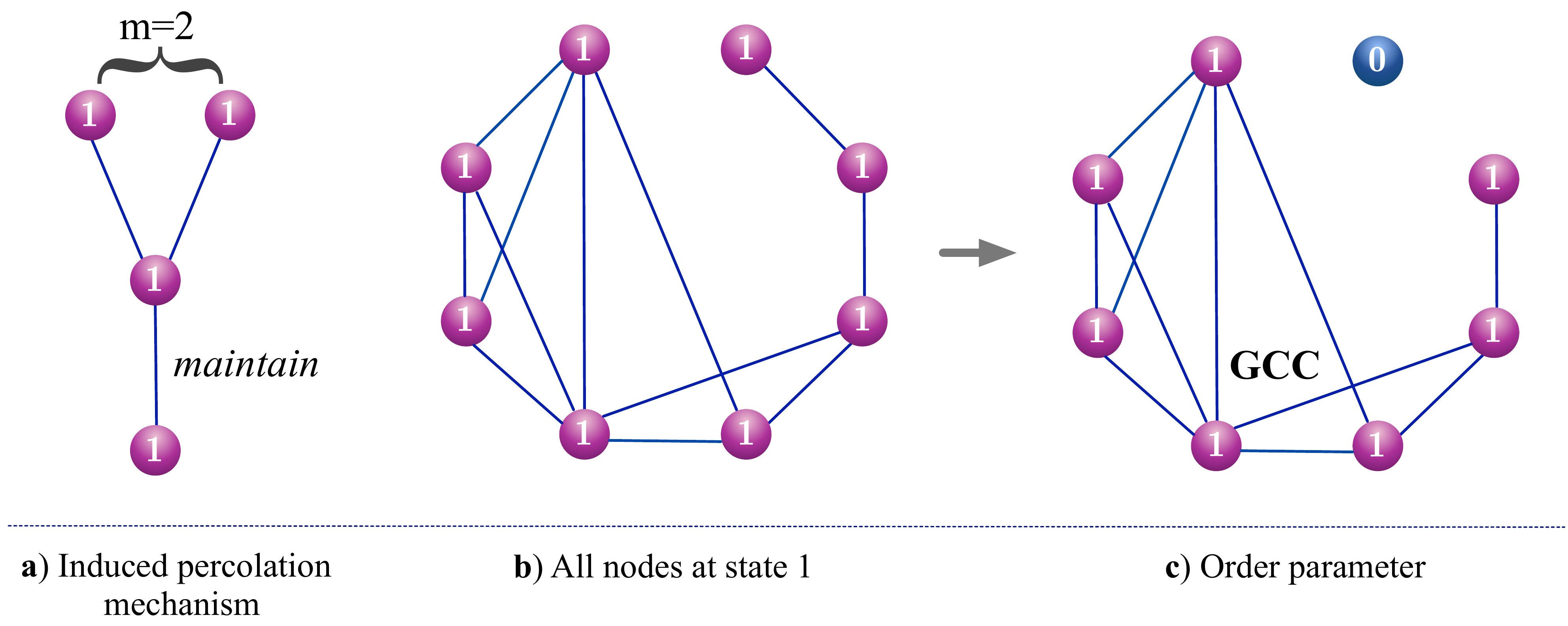}%
	\caption{Induced percolation on undirected networks.}%
	\label{fig_inducing_undirected}%
\end{figure}

\begin{figure}[!htp]
	\centering
	\includegraphics[width=.8\textwidth]{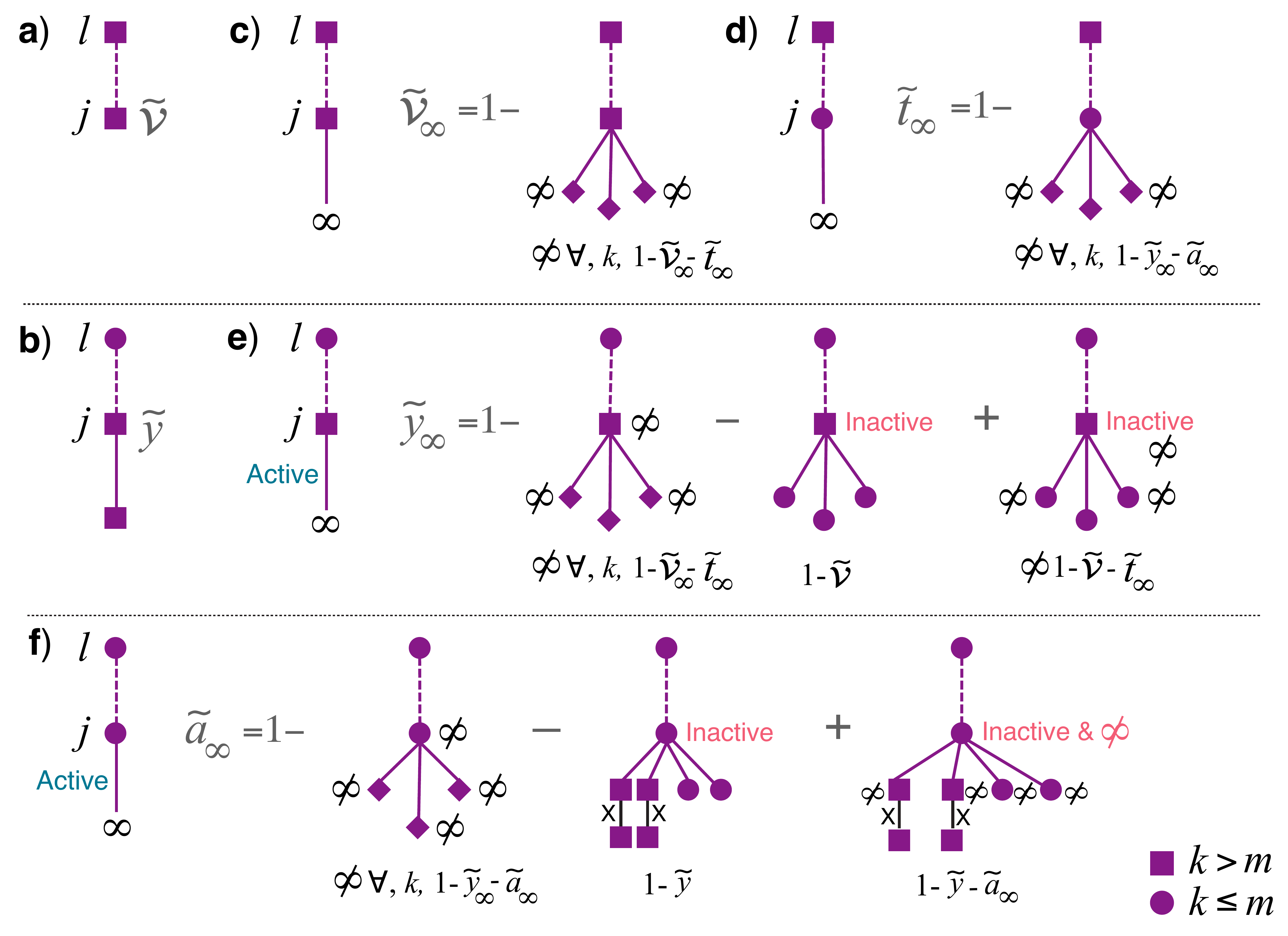}
	\caption{Graphical representation of how the conditional probabilities are calculated for undirected networks.}
	\label{fig_uninduced_notation}%
\end{figure}

We presented the theoretical analysis of induced percolation on undirected networks in the Methods section of the main text. Here, we supplement the analysis with illustrations on the definition of induced percolation on undirected networks (as shown in Figure \ref{fig_inducing_undirected}). In addition, we illustrate the relation between conditional probabilities (as shown in Figure \ref{fig_uninduced_notation}) defined when deriving the order parameter $ P_\infty $ on undirected networks. 

\subsection{Relation between different order parameters in directed networks}
In directed networks, there are three types of giant connected components: giant strongly connected component (GSCC), giant out-going component (GOUT) and giant in-coming component (GIN), as shown in Figure 1 in the main text. Within a giant strongly connected component, any two nodes can reach each other through directed links. The GSCC corresponds to the largest strongly connected component above the critical point. GOUT corresponds to the set of active nodes that can be reached along directed links starting from the nodes in the GSCC, while GIN corresponds to the set of active nodes that can reach the GSCC along directed links. Therefore, GSCC is a subset of GOUT and GIN. In undirected networks, GSCC, GOUT, and GIN are the same set of nodes.


	\begin{table*}
	\caption{
		Characteristics of the $5$ real-world networks used in Fig. 5 of the main text. Columns from the second to the fifth present the number $ N $ of nodes, the number $ E $ of links, network type and the average degree $\langle k \rangle = \frac{2E}{N}$.} \label{tab:real_networks}
	\begin{tabular}{cccccc}
		\hline
		Network & $N$ & $E$ & Network type & $\langle k \rangle$ \\ \toprule
		P2p-04 \cite{Ripeanu.Iamnitchi.Foster-IEEEInternetComputing-2002, Leskovec.Kleinberg.Faloutsos-TransKDD-2007} & 10,876 & 39,994 & Directed & 7.35 \\ \hline
		P2p-24 \cite{Ripeanu.Iamnitchi.Foster-IEEEInternetComputing-2002, Leskovec.Kleinberg.Faloutsos-TransKDD-2007} & 26,518 & 65,369 & Directed & 4.93 \\ \hline
		Email-Enron \cite{leskovec2009community,klimt2004introducing} & 36,692 & 183,831 & Undirected & 10.02 \\ \hline
		Collab-HepTh \cite{Leskovec.Kleinberg.Faloutsos-TransKDD-2007} & 9,875 & 25,973 & Undirected & 5.26 \\ \hline
		Collab-GrQc \cite{Leskovec.Kleinberg.Faloutsos-TransKDD-2007} & 5,241 & 14,484 & Undirected & 5.53 \\ \hline
	\end{tabular}
\end{table*}

\section*{Acknowledgments}
X.W. is supported by the National Natural Science Foundation of China under grant No. 62003156 and by ``PCL Future Greater-Bay Area Network Facilities for Large-scale Experiments and Applications (LZC0019)". Y.M. acknowledges partial support from the Government of Arag\'on, Spain through a grant to the group FENOL (E36-20R), by MINECO and FEDER funds (grant FIS2017-87519-P) and by Intesa Sanpaolo Innovation Center.

%
\end{document}